\documentclass[english,aps,pre,preprint,nofootinbib]{revtex4-1}
\usepackage[T1]{fontenc}
\usepackage[latin9]{inputenc}
\usepackage{units}
\usepackage{bm}
\usepackage{amsmath}
\usepackage{mathdots}
\usepackage{stmaryrd}
\usepackage{graphicx}
\usepackage{xargs}[2008/03/08]

\makeatletter
\usepackage{bm}
\usepackage{color}

\newcommand{\C}[1]{{}}
\newcommand{\bS}{\begin{subequations}}
\newcommand{\eS}{\end{subequations}}

\DeclareMathOperator{\Pf}{Pf}
\DeclareMathOperator{\Tr}{Tr}
\DeclareMathOperator{\diag}{\mathbf{diag}}
\DeclareMathOperator{\sign}{sign}
\DeclareMathOperator{\QP}{\Pi}

\newcommand{\para}{{\mkern2mu\vphantom{\perp}\vrule depth 0pt\mkern2mu\vrule depth 0pt\mkern2mu}}

\makeatother

\usepackage{babel}
\begin{document}
\global\long\def\mat#1{\bm{\mathrm{#1}}}

\C{\global\long\def\Pf{\mathrm{Pf}}
\global\long\def\para{\parallel}
\global\long\def\Tr{\mathrm{Tr}}
\global\long\def\diag{\mat{\mathrm{diag}}}
\global\long\def\sign{\mathrm{sign}}
\global\long\def\QP{\Pi}
 see preamble }

\global\long\def\Def{\mathcal{\equiv}}
\let\Def=\equiv

\global\long\def\T#1{\bar{#1}}
\global\long\def\Tc{T_{\mathrm{c}}}

\global\long\def\zc{z_{\mathrm{c}}}
\global\long\def\ii{\mathrm{i}}
\global\long\def\ee{\mathrm{e}}
\global\long\def\dd{\mathrm{d}}

\global\long\def\L{\leftrightarrow}
\global\long\def\M{\updownarrow}

\global\long\def\Mh{\nicefrac{M}{2}}
\global\long\def\Det{\mathcal{Z}}
\global\long\def\TM#1{\mat{\mathcal{T}}_{\!\!#1}}

\newcommandx\Vt[1][usedefault, addprefix=\global, 1=]{\mat{\mathcal{V}}_{#1}^{\M}}
\newcommandx\Vz[1][usedefault, addprefix=\global, 1=]{\mat{\mathcal{V}}_{#1}^{\L}}

\global\long\def\S{\vphantom{\T{\mathbf{s}}}\mathbf{s}}
\global\long\def\Sc{\T{\mathbf{s}}}
\C{\global\long\def\So{\vphantom{\bar{\mathrm{o}}}\mathrm{o}}
\global\long\def\Soc{\bar{\mathrm{o}}}
}\global\long\def\So{\mathbf{o}}
\global\long\def\Soc{\mathbf{e}}

\global\long\def\inf{\infty}
\global\long\def\strip{\mathrm{strip}}

\global\long\def\Zstrip{Z_{\strip}}
\global\long\def\Fstrip{F_{\strip}}

\global\long\def\Zinf{Z_{\inf}}
\global\long\def\Finf{F_{\inf}}

\global\long\def\Zres{Z_{\inf}^{\mathrm{res}}}
\global\long\def\Fres{F_{\inf}^{\mathrm{res}}}

\global\long\def\Zsres{Z_{\strip}^{\mathrm{res}}}
\global\long\def\Fsres{F_{\strip}^{\mathrm{res}}}

\global\long\def\fb{f_{\mathrm{b}}}
\global\long\def\fs{f_{\mathrm{s}}}
\global\long\def\fc{f_{\mathrm{c}}}
\global\long\def\FC{\mathcal{F}}

\global\long\def\dom#1{\hat{#1}}

\renewcommand*{\arraystretch}{0.8}

\title{The square lattice Ising model on the rectangle\\
 I: Finite systems}

\author{Alfred Hucht}

\affiliation{Faculty for Physics, University of Duisburg-Essen, 47058 Duisburg,
Germany}

\date{\today}
\begin{abstract}
The partition function of the square lattice Ising model on the rectangle
with open boundary conditions in both directions is calculated exactly
for arbitrary system size $L\times M$ and temperature. We start with
the dimer method of Kasteleyn, McCoy \& Wu, construct a highly symmetric
block transfer matrix and derive a factorization of the involved determinant,
effectively decomposing the free energy of the system into two parts,
$F(L,M)=\Fstrip(L,M)+\Fsres(L,M)$, where the residual part $\Fsres(L,M)$
contains the nontrivial finite-$L$ contributions for fixed $M$.
It is given by the determinant of a $\Mh\times\Mh$ matrix and can
be mapped onto an effective spin model with $M$ Ising spins and long-range
interactions. While $\Fsres(L,M)$ becomes exponentially small for
large $L/M$ or off-critical temperatures, it leads to important finite-size
effects such as the critical Casimir force near criticality. The relations
to the Casimir potential and the Casimir force are discussed.
\end{abstract}
\maketitle
\tableofcontents{}

\vfill{}

\pagebreak{}

\section{Introduction }

The two-dimensional Ising model \cite{Ising25} on the $L\times M$
square lattice is one of the best investigated models in statistical
mechanics. After the exact solution of the periodic case by Onsager
\cite{Onsager44}, many authors have contributed to the knowledge
about this model under various aspects, such as different boundary
conditions (BCs) or surface effects \cite{McCoyWu73,Baxter82}. Near
the critical temperature $\Tc$, where the correlation length $\xi(T)$
of thermal fluctuations becomes of the order of the system size $L$
or $M$ in finite systems, interesting finite-size effects such as
the critical Casimir effect emerge, which describes an interaction
of the system boundaries mediated by long-range critical fluctuations
\cite{FisherdeGennes78} in close analogy to the quantum electrodynamical
Casimir effect \cite{Casimir48}. These finite-size effects can be
described by universal finite-size scaling functions, that only depend
on the bulk and surface universality classes of the model, as well
as on the BCs and on the system shape. They have been calculated exactly
for many cases, albeit mostly in strip geometry, where the aspect
ratio $\rho=L/M$ of the system goes to zero \cite{EvansStecki94,Au-YangFisher80,BrankovDantchevTonchev00}.
Directly at the critical point, exact methods or conformal field theory
can be used to get exact expressions for the Casimir amplitude $\Delta_{\mathrm{C}}(\rho)$
for arbitrary $\rho$. This has been done for periodic \cite{FerdinandFisher69,LuWu01}
as well as for open BCs \cite{KlebanVassileva91}. At arbitrary aspect
ratios and temperatures, however, the finite-size scaling functions
must be derived from the exact solution of the system with the correct
BC. For the Ising model, this has been done only in a few cases, namely
for the torus with periodic BCs in both directions \cite{HuchtGruenebergSchmidt11,HobrechtHucht16a}
and for the cylinder with open BCs in one direction \cite{HobrechtHucht16a}. 

In this work and in the forthcoming publication \cite{Hucht16b} we
will present a calculation of these finite-size contributions, namely
the residual free energy also denoted Casimir potential, as well as
the resulting critical Casimir forces, for open BCs at arbitrary temperatures
$T$ and system size $L\times M$. In order to calculate these quantities
correctly, all infinite volume free energies, i.\,e., the bulk free
energy $LM\fb(T)$, the surface free energies $L\fs^{\L}(T)$ and
$M\fs^{\M}(T)$ in the two directions $\L$ and $\M$, as well as
the corner free energy $\fc(T)$ must be known and subtracted from
the free energy of the finite system. While the bulk and surface free
energies are known for a long time \cite{Onsager44,McCoyWu73}, the
corner free energy $\fc(T)$ was only known below $\Tc$ from a conjecture
by Vernier \& Jacobsen \cite{VernierJacobsen12}. The corresponding
product formula for the paramagnetic phase is given in the Appendix
of this work and will be discussed in \cite{Hucht16b}. 

In a recent preprint \cite{Baxter16}, Rodney J.~Baxter presented
an exact calculation of the infinite volume corner free energy $\fc(T)$
in the ordered phase $T<\Tc$, verifying the conjecture of Vernier
\& Jacobsen. In this manuscript we present a calculation within the
same model and geometry and discuss the similarities and differences.
While Baxter focused on the corner free energy contribution $\fc(T)$
in the thermodynamic limit, the focus of this work is on the exact
finite-size corrections to the free energy at arbitrary system size
and temperature. 

We start the present calculation with the Pfaffian formulation of
Kasteleyn, McCoy \& Wu \cite{Kasteleyn63,McCoyWu73} in cylinder geometry
and reduce the involved determinant of a sparse $4LM\times4LM$ matrix
to the determinant of a $LM\times LM$ block-tridiagonal matrix using
an appropriate Schur complement. This determinant can then be calculated
with the formula of Molinari \cite{Molinari08}, introducing $2\times2$
block transfer matrices $\mat{\mathcal{T}}_{\negmedspace\ell}$ with
$M\times M$ blocks. Up to here the calculation is done for arbitrary
local couplings $K_{\ell,m}^{\L}$ and $K_{\ell,m}^{\M}$ in the two
directions on the cylinder. Then we assume open BCs in both directions
and homogeneous, albeit anisotropic couplings $K^{\L}$ and $K^{\M}$.
After that simplification the partition function $Z$ is of the form
$Z^{2}\propto\det\langle\mat 1\,\mat 0|\mat{\mathcal{T}}^{L}|\mat 1\,\mat 0\rangle$,
in strong analogy to Baxter's result \cite{Baxter16}.

While Baxter at this point performs the thermodynamic limit $L\to\infty$
with fixed $M$, neglecting the finite-$L$ contributions, we are
able to proceed and further reduce the size of the involved matrices.
The block transfer matrix $\mat{\mathcal{T}}$ can be symmetrized
and block diagonalized such that its eigenvalues $\lambda$ are real
and occur in pairs $(\lambda,\lambda^{-1})$, and the calculation
is simplified by the introduction of the natural angle variable $\varphi$,
leading to the characteristic polynomial $P_{M}(\varphi)$. It turns
out that the eigenvalues $\lambda$ are directly related to the well-known
Onsager-$\gamma$ via $\gamma=\log\lambda$.

The eigenvectors $\vec{X}$ of $\mat{\mathcal{T}}$ show an important
symmetry with respect to the mapping $\lambda\leftrightarrow\lambda^{-1}$,
which can eventually be used to reduce the involved matrices from
$2M\times2M$ to $M\times M$ and, more important, to factorize the
determinant into a product of the form $\det(\mat W^{T}\mat D\mat W)=\det^{2}\mat W\det\mat D$,
where $\mat D$ is diagonal.

The remaining matrix $\mat W$ is of Vandermonde type and can be considerably
simplified using the invariance property of Vandermonde determinants
with respect to basis transformations. Using the well known product
formula for these determinants the matrix size can be further reduced
to $\Mh\times\Mh$. We show that this determinant contains all remaining
nontrivial finite-size contributions, and discuss the different resulting
contributions to the free energy.

Finally we present an exact mapping of the remaining determinant onto
a long-range spin model with $M$ spins and logarithmic interactions
in an effective magnetic field of strength $L$, which might give
rise to an alternative calculation of the remaining determinant. We
conclude with a discussion of the results. 

In the second part of this work \cite{Hucht16b}, which will be published
separately, we perform the finite-size scaling limit $L,M\to\infty$,
$T\to\Tc$ with fixed temperature scaling variable $x\propto(T/\Tc-1)M$
and fixed aspect ratio $\rho$. After a number of simplifications,
we derive exponentially fast converging series for the Casimir scaling
functions. At the critical point $T=\Tc$ we can rewrite the Casimir
amplitude $\Delta_{\mathrm{C}}(\rho)$ in terms of the Dedekind eta
function, confirming a prediction from conformal field theory \cite{KlebanVassileva91}.

\section{Model and Pfaffian representation}

\begin{figure}
\begin{centering}
\includegraphics[scale=0.45]{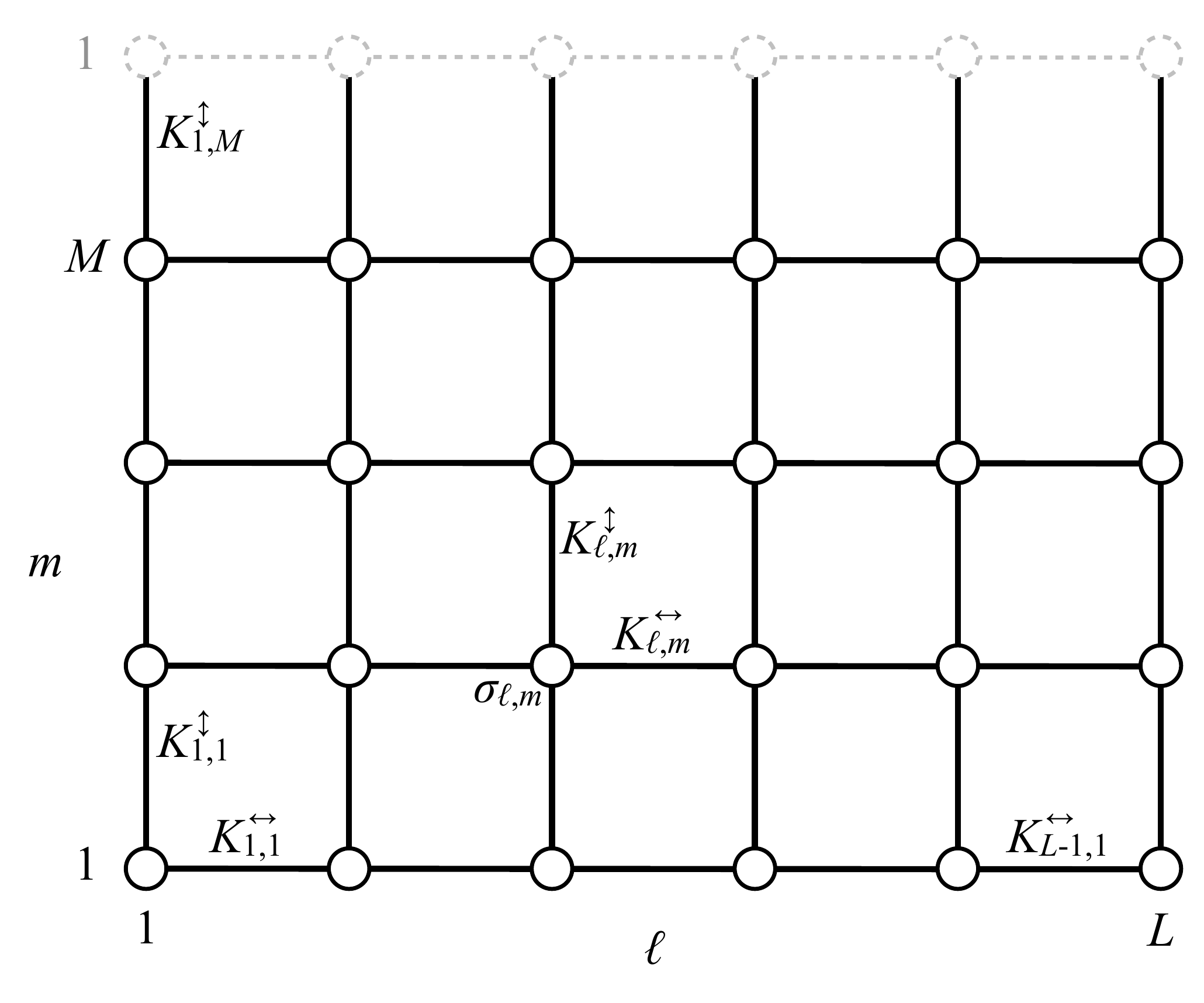}
\par\end{centering}
\caption{The square lattice with cylinder geometry for $M=4$ and $L=6$.\label{fig:lattice}}
\end{figure}
We consider the Ising model on the square lattice with $L$ columns
and $M$ rows as shown in figure~\ref{fig:lattice}, and start with
arbitrary reduced (in units of $k_{\mathrm{B}}T$, with Boltzmann
constant $k_{\mathrm{B}}$) couplings $K_{\ell,m}^{\L}$ and $K_{\ell,m}^{\M}$
in horizontal and vertical direction on the cylinder periodic in vertical
($M$) direction. Our aim is to calculate the partition function 
\begin{equation}
Z=\Tr\exp\sum_{\ell=1}^{L}\sum_{m=1}^{M}\left(K_{\ell,m}^{\L}\sigma_{\ell,m}\sigma_{\ell+1,m}+K{}_{\ell,m}^{\M}\sigma_{\ell,m}\sigma_{\ell,m+1}\right),\label{eq:Ising}
\end{equation}
where the trace is over all $2^{LM}$ configurations of the $LM$
spins $\sigma_{\ell,m}=\pm1$, with $\sigma_{L+1,m}=0$ and $\sigma_{\ell,M+1}=\sigma_{\ell,1}$.
We assume open BC in horizontal ($L$) direction, $K_{L,m}^{\L}=0$,
and first derive a transfer matrix formulation for this general case.
After that we focus on the rectangular homogeneous case, $K_{\ell,M}^{\M}=0$,
$K_{\ell,m<M}^{\M}=K^{\M}$, $K_{\ell<L,m}^{\L}=K^{\L}$, where we
still allow for anisotropic couplings.

Our starting point is the Pfaffian representation by Kasteleyn, McCoy
\& Wu \cite{Kasteleyn63,McCoyWu73}, where the partition function
in cylinder geometry is given by 
\begin{equation}
Z=\sqrt{C_{0}}\,\Pf\mat{\mathcal{A}}=\sqrt{C_{0}\det\mat{\mathcal{A}}}\,,\label{eq:Z_a}
\end{equation}
with the constant 
\begin{align}
C_{0} & \Def4^{LM}\prod_{\ell=1}^{L-1}\prod_{m=1}^{M}\cosh^{2}K_{\ell,m}^{\L}\;\prod_{\ell=1}^{L}\prod_{m=1}^{M}\cosh^{2}K_{\ell,m}^{\M}.\label{eq:K_0}
\end{align}
We define the antisymmetric $4LM\times4LM$ sparse matrix $\mat{\mathcal{A}}$
as a $4\times4$ block matrix (the bar denotes transposition, ``$\Def$''
denotes a definition)
\begin{equation}
\mat{\mathcal{A}}\Def\begin{bmatrix}\mat 0 & \mat 1+\mat Z^{\M} & -\mat 1 & -\mat 1\\
-\mat 1-\T{\mat Z}^{\M} & \mat 0 & \mat 1 & -\mat 1\\
\mat 1 & -\mat 1 & \mat 0 & \mat 1+\mat Z^{\L}\\
\mat 1 & \mat 1 & -\mat 1-\T{\mat Z}^{\L} & \mat 0
\end{bmatrix},\label{eq:A4x4}
\end{equation}
where the $LM\times LM$ matrices $\mat Z^{\delta}$ contain the couplings
$z_{\ell,m}^{\delta}\Def\tanh K_{\ell,m}^{\delta}$ in direction $\delta=\,\L,\M$
via the $M\times M$ and $LM\times LM$ diagonal matrices 
\begin{equation}
\mat z_{\ell}^{\delta}\Def\diag(z_{\ell,1}^{\delta},\ldots,z_{\ell,M}^{\delta}),\quad\mat z^{\delta}\Def\diag(\mat z_{1}^{\delta},\ldots,\mat z_{L}^{\delta}),\label{eq:zDef}
\end{equation}
according to \bS\label{eq:Z_LM}
\begin{align}
\mat Z^{\L} & \Def\mat z^{\L}(\mat H_{L}^{0}\otimes\mat 1_{M}),\label{eq:Z_L}\\
\mat Z^{\M} & \Def\mat z^{\M}(\mat 1_{L}\otimes\mat H_{M}^{-})=\diag(\mat z_{1}^{\M}\mat H_{M}^{-},\ldots,\mat z_{L}^{\M}\mat H_{M}^{-}).\label{eq:Z_M}
\end{align}
\eS{}Here we have introduced the $n\times n$ shift matrices
\begin{equation}
\setlength{\arraycolsep}{8pt}\mat H_{n}^{0}\Def\begin{pmatrix}0 & 1\\
 & \ddots & \ddots\\
 &  & \ddots & 1\\
0 &  &  & 0
\end{pmatrix},\quad\mat H_{n}^{-}\Def\begin{pmatrix}0 & 1\\
 & \ddots & \ddots\\
 &  & \ddots & 1\\
-1 &  &  & 0
\end{pmatrix},\label{eq:H_shift}
\end{equation}
that, together with the $n\times n$ identity matrix $\mat 1_{n}$,
define the lattice structure. We drop the index $n$ from unit and
zero matrices $\mat 1$, $\mat 0$ as long as it can be implied from
the context.

\section{Schur reduction}

\global\long\def\idx#1{#1}
We first reduce the matrix size from $4LM\times4LM$ to $LM\times LM$
by a standard Schur reduction according to 
\begin{equation}
\det\mat{\mathcal{A}}=\det\mat{\mathcal{A}}_{\idx{\bar{i},\bar{i}}}\det\mat{\mathcal{C}}_{\idx{i,i}},\label{eq:detA_reduction}
\end{equation}
where $\bar{i}$ denotes the index complement of $i$, i.\,e.~$\mat{\mathcal{A}}_{\idx{\bar{i},j}}$
is derived from $\mat{\mathcal{A}}$ by dropping row $i$ and taking
column $j$. We choose $i=4$ to find, for even $M$, 
\begin{equation}
\det\mat{\mathcal{A}}_{\idx{\bar{4},\bar{4}}}=\prod_{\ell=1}^{L}\Bigg(\prod_{{m=1\atop m\,\mathrm{odd}}}^{M-1}z_{\ell,m}^{\M}+\prod_{{m=2\atop m\,\mathrm{even}}}^{M}z_{\ell,m}^{\M}\Bigg)^{2}\label{detA_4c4c}
\end{equation}
as well as the $LM\times LM$ Schur complement 
\begin{equation}
\mat{\mathcal{C}}_{\idx{4,4}}\Def\mat{\mathcal{A}}/\mat{\mathcal{A}}_{\idx{\bar{4},\bar{4}}}\Def\mat{\mathcal{A}}_{\idx{4,4}}^{\vphantom{-1}}-\mat{\mathcal{A}}_{\idx{4,\bar{4}}}^{\vphantom{-1}}\mat{\mathcal{A}}_{\idx{\bar{4},\bar{4}}}^{-1}\mat{\mathcal{A}}_{\idx{\bar{4},4}}^{\vphantom{-1}},\label{eq:C_44-def}
\end{equation}
which is antisymmetric and block tridiagonal,
\begin{equation}
\mat{\mathcal{C}}_{\idx{4,4}}=\begin{bmatrix}\mat A_{1} & \mat B_{1}\\
-\T{\mat B}_{1} & \ddots & \ddots\\
 & \ddots & \ddots & \mat B_{L-1}\\
 &  & -\T{\mat B}_{L-1} & \mat A_{L}
\end{bmatrix},\label{eq:C_44}
\end{equation}
with $M\times M$ matrices $\mat A_{\ell}$ and $\mat B_{\ell}$.
We also could have chosen $i=3$ for the reduction, which would reflect
the matrix $\mat{\mathcal{C}}_{\idx{i,i}}$ along the anti-diagonal,
whereas the indices $i=1,2$ do not lead to block tridiagonal matrices
$\mat{\mathcal{C}}_{\idx{i,i}}$. The explicit expressions for the
matrices $\mat A_{\ell}$ and $\mat B_{\ell}$ are \bS
\begin{align}
\mat B_{\ell}^{-1} & =-(\mat z_{\ell}^{\L})^{-1}\mat D_{\ell},\label{eq:B_ell}\\
\mat A_{1} & =\mat A_{1}^{-},\label{eq:A_1}\\
\mat A_{\ell>1} & =\mat A_{\ell}^{-}+\mat z_{\ell-1}^{\L}\mat A_{\ell-1}^{+}\mat z_{\ell-1}^{\L},\label{eq:A_ell}
\end{align}
\eS{}with the auxiliary matrices \bS 
\begin{align}
\mat A_{\ell}^{\pm} & \Def\pm\left[(\mat 1\pm\T{\mat Z}_{\ell}^{\M})^{-1}-(\mat 1\pm\mat Z_{\ell}^{\M})^{-1}\right]^{-1},\label{eq:Apm_ell}\\
\mat D_{\ell} & \Def(\mat 1-\T{\mat Z}_{\ell}^{\M})(\mat 1-\mat Z_{\ell}^{\M}\T{\mat Z}_{\ell}^{\M})^{-1}-(\mat 1-\mat Z_{\ell}^{\M})(\mat 1-\T{\mat Z}_{\ell}^{\M}\mat Z_{\ell}^{\M})^{-1},\label{eq:D_ell}
\end{align}
where $\mat Z_{\ell}^{\M}=\mat z_{\ell}^{\M}\mat H_{M}^{-}$ from
(\ref{eq:Z_M}). As the matrices $\mat B_{\ell}$ are invertible,
the remaining determinant $\det\mat{\mathcal{C}}_{\idx{4,4}}$ can
be calculated with a transfer matrix approach.

\section{The block transfer matrix $\protect\mat{\mathcal{T}}$}

\eS{}The determinant of the block tridiagonal matrix $\mat{\mathcal{C}}_{\idx{4,4}}$
from (\ref{eq:C_44}) can be calculated with the method of Molinari
\cite{Molinari08}. We introduce the $2\times2$ block transfer matrix
(TM)
\begin{equation}
\mat{\mathcal{T}}_{\ell,\ell-1}^{\ddagger}\Def\begin{bmatrix}-\mat B_{\ell}^{-1}\mat A_{\ell} & \mat B_{\ell}^{-1}\T{\mat B}_{\ell-1}\\
\mat 1 & \mat 0
\end{bmatrix},\label{eq:T''_a}
\end{equation}
with $\ell=1,\ldots,L$, and formally define $\mat B_{0}$ and $\mat B_{L}$,
with $\mat z_{0}^{\delta}=\mat z_{L}^{\M}=\mat 0$ and $\mat z_{L}^{\L}=\mat 1$,
in order to keep the expressions simple. We can factorize $\mat{\mathcal{T}}_{\ell,\ell-1}^{\ddagger}$
into two parts depending on $\ell$ and $\ell-1$, respectively, 
\begin{equation}
\mat{\mathcal{T}}_{\ell,\ell-1}^{\ddagger}=\begin{bmatrix}(\mat z_{\ell}^{\L})^{-1}\mat D_{\ell}\mat A_{\ell}^{-} & (\mat z_{\ell}^{\L})^{-1}\mat D_{\ell}\\
\mat 1 & \mat 0
\end{bmatrix}\begin{bmatrix}\mat 1 & \mat 0\\
\mat z_{\ell-1}^{\L}\mat A_{\ell-1}^{+}\mat z_{\ell-1}^{\L} & \mat z_{\ell-1}^{\L}\T{\mat D}_{\ell-1}^{-1}
\end{bmatrix}\Def\mat{\mathcal{T}}_{\ell}^{(1)}\mat{\mathcal{T}}_{\ell-1}^{(2)},\label{eq:T''_b}
\end{equation}
and we observe that in the product of TMs, $\cdots\mat{\mathcal{T}}_{\ell+1,\ell}^{\ddagger}\mat{\mathcal{T}}_{\ell,\ell-1}^{\ddagger}\cdots=\cdots\mat{\mathcal{T}}_{\ell+1}^{(1)}\mat{\mathcal{T}}_{\ell}^{(2)}\mat{\mathcal{T}}_{\ell}^{(1)}\mat{\mathcal{T}}_{\ell-1}^{(2)}\cdots$,
we can identify a shifted TM $\mat{\mathcal{T}}_{\ell}^{\dagger}\Def\mat{\mathcal{T}}_{\ell}^{(2)}\mat{\mathcal{T}}_{\ell}^{(1)}$,
depending only on $\ell$, with the factorization 
\begin{equation}
\mat{\mathcal{T}}_{\ell}^{\dagger}\Def\mat{\mathcal{T}}_{\ell}^{(2)}\mat{\mathcal{T}}_{\ell}^{(1)}=\begin{bmatrix}(\mat z_{\ell}^{\L})^{-1} & \mat 0\\
\mat 0 & \mat z_{\ell}^{\L}
\end{bmatrix}\begin{bmatrix}\mat 1 & \mat 0\\
\mat A_{\ell}^{+} & \mat 1
\end{bmatrix}\begin{bmatrix}\mat 0 & \mat D_{\ell}\\
\T{\mat D}_{\ell}^{-1} & \mat 0
\end{bmatrix}\begin{bmatrix}\mat 1 & \mat 0\\
\mat A_{\ell}^{-} & \mat 1
\end{bmatrix}.\label{eq:T'_a}
\end{equation}
Using a block rotation by $\theta=\pi/4$, with 
\begin{equation}
\mat R_{\theta}\Def\mat r_{\theta}\otimes\mat 1,\qquad\mat r_{\theta}\Def\begin{pmatrix}\cos\theta & \sin\theta\\
-\sin\theta & \cos\theta
\end{pmatrix},\label{eq:Rtheta}
\end{equation}
we find the simple representation
\begin{equation}
\T{\mat R}_{\frac{\pi}{4}}\begin{bmatrix}\mat 1 & \mat 0\\
\mat A_{\ell}^{+} & \mat 1
\end{bmatrix}\begin{bmatrix}\mat 0 & \mat D_{\ell}\\
\T{\mat D}_{\ell}^{-1} & \mat 0
\end{bmatrix}\begin{bmatrix}\mat 1 & \mat 0\\
\mat A_{\ell}^{-} & \mat 1
\end{bmatrix}\mat R_{\frac{\pi}{4}}=\begin{bmatrix}\T{\mat H}^{-} & \mat 0\\
\mat 0 & \mat 1
\end{bmatrix}\begin{bmatrix}\mat t_{\ell,+} & \mat t_{\ell,-}\\
\mat t_{\ell,-} & \mat t_{\ell,+}
\end{bmatrix}\begin{bmatrix}\mat H^{-} & \mat 0\\
\mat 0 & \mat 1
\end{bmatrix}\Def\Vt[\ell],\label{eq:V_t,ell}
\end{equation}
where the matrices
\begin{equation}
\mat t_{\ell}\Def\diag(t_{\ell,1},\ldots,t_{\ell,M})\label{eq:t_ell}
\end{equation}
contain the dual couplings $t\Def z^{\M\,*}=\frac{1-z^{\M}}{1+z^{\M}}$
of $z^{\M}$. We have introduced the abbreviation 
\begin{equation}
a_{\pm}\Def{\textstyle \frac{1}{2}}(a\pm a^{-1}),\label{eq:apm}
\end{equation}
such that $a^{\pm1}=a_{+}\pm a_{-}$, for couplings and other quantities.
From here on we express the vertical couplings $z^{\M}$ through their
dual couplings $t$, and simply write $z$ for the horizontal couplings
$z^{\L}$. Note that our $z$ is denoted $u$ in \cite{Baxter16}. 

Inserting three $\mat 1$s into (\ref{eq:T'_a}), we find 
\begin{align}
\mat{\mathcal{T}}_{\ell}^{\dagger} & =\mat R_{\frac{\pi}{4}}\underbrace{\T{\mat R}_{\frac{\pi}{4}}\begin{bmatrix}\mat z_{\ell}^{-1} & \mat 0\\
\mat 0 & \mat z_{\ell}
\end{bmatrix}\mat R_{\frac{\pi}{4}}}_{\Vz[\ell]}\underbrace{\T{\mat R}_{\frac{\pi}{4}}\begin{bmatrix}\mat 1 & \mat 0\\
\mat A_{\ell}^{+} & \mat 1
\end{bmatrix}\begin{bmatrix}\mat 0 & \mat D_{\ell}\\
\T{\mat D}_{\ell}^{-1} & \mat 0
\end{bmatrix}\begin{bmatrix}\mat 1 & \mat 0\\
\mat A_{\ell}^{-} & \mat 1
\end{bmatrix}\mat R_{\frac{\pi}{4}}}_{\Vt[\ell]}\T{\mat R}_{\frac{\pi}{4}}\nonumber \\
 & =\mat R_{\frac{\pi}{4}}\Vz[\ell]\Vt[\ell]\T{\mat R}_{\frac{\pi}{4}},\label{eq:T'_b}
\end{align}
with
\begin{equation}
\Vz[\ell]=\begin{bmatrix}\mat z_{\ell,+} & -\mat z_{\ell,-}\\
-\mat z_{\ell,-} & \mat z_{\ell,+}
\end{bmatrix}\label{eq:V_z,ell}
\end{equation}
in analogy to equation~(\ref{eq:V_t,ell}). Following \cite{Molinari08},
the determinant (\ref{eq:detA_reduction}) becomes
\begin{align}
\det\mat{\mathcal{A}} & =C_{1}\det\langle\mat 1\,\mat 0|\mat{\mathcal{T}}_{L,L-1}^{\ddagger}\mat{\mathcal{T}}_{L-1,L-2}^{\ddagger}\cdots\mat{\mathcal{T}}_{2,1}^{\ddagger}\mat{\mathcal{T}}_{1,0}^{\ddagger}|\mat 1\,\mat 0\rangle\nonumber \\
 & =C_{1}\det\langle\mat 1\,\mat 0|\mat{\mathcal{T}}_{L}^{\dagger}\mat{\mathcal{T}}_{L-1}^{\dagger}\cdots\mat{\mathcal{T}}_{2}^{\dagger}\mat{\mathcal{T}}_{1}^{\dagger}|\mat 1\,\mat 0\rangle\nonumber \\
 & =C_{1}\det\langle\mat e|\Vz[L]\Vt[L]\,\Vz[L-1]\Vt[L-1]\cdots\Vz[2]\Vt[2]\,\Vz[1]\Vt[1]|\mat e\rangle,\label{eq:detA}
\end{align}
with $|\mat e\rangle\Def\T{\mat R}_{\frac{\pi}{4}}|\mat 1\,\mat 0\rangle=\frac{1}{\sqrt{2}}|\mat 1\,\mat 1\rangle$
and the constant
\begin{equation}
C_{1}\Def\det\mat{\mathcal{A}}_{\idx{\bar{4},\bar{4}}}\prod_{\ell=1}^{L}\det\mat B_{\ell}=\prod_{\ell=1}^{L-1}\prod_{m=1}^{M}z_{\ell,m}^{\L}\;\prod_{\ell=1}^{L}\prod_{m=1}^{M}\big(1-z_{\ell,m}^{\M\,2}\big).\label{eq:K_1}
\end{equation}
Here and in the following we use bra-ket notation for the boundary
block vectors, such that $\langle\mat e|$ and $|\mat e\rangle$ are
$M\times2M$ and $2M\times M$ dimensional matrices, respectively,
and $\langle\mat 1\,\mat 0|\TM{}|\mat 1\,\mat 0\rangle$ gives the
$1{,}1$-element of block matrix $\TM{}$. 

The final result for the partition function (\ref{eq:Z_a}) with arbitrary
couplings reads \bS \label{eq:Z_arbitrary}
\begin{equation}
Z=\sqrt{C_{2}^{\dagger}\Det_{\vphantom{2}}^{\dagger}},\label{eq:Z_b}
\end{equation}
with
\begin{equation}
\Det^{\dagger}\Def\det\langle\mat e|\Vt[L]\,\Vz[L-1]\Vt[L-1]\cdots\Vz[2]\Vt[2]\,\Vz[1]\Vt[1]|\mat e\rangle,\label{eq:Z_Det}
\end{equation}
as $\Vz[L]=\mat 1$, and with the constant
\begin{equation}
C_{2}^{\dagger}\Def C_{0}C_{1}=2^{(L+1)M}\prod_{\ell=1}^{L-1}\prod_{m=1}^{M}\frac{1}{z_{\ell,m,-}}.\label{eq:K_2}
\end{equation}
\eS{}This result is valid for arbitrary couplings on the cylinder,
and it is straightforward to derive an analog expression for the torus.
We point out that we can ``transpose'' both $\Vz[\ell]$ and $\Vt[\ell]$
from $2\times2$ block structure with $M\times M$ blocks to $M\times M$
block structure with $2\times2$ blocks to get, for $M=4$, \begingroup\renewcommand*{\arraystretch}{0.8}\bS
\begin{align}
\hat{\mat{\mathcal{V}}}_{\ell}^{\leftrightarrow} & =\begin{pmatrix}z_{\ell,1,+} & -z_{\ell,1,-}\\
-z_{\ell,1,-} & z_{\ell,1,+}\\
 &  & z_{\ell,2,+} & -z_{\ell,2,-}\\
 &  & -z_{\ell,2,-} & z_{\ell,2,+}\\
 &  &  &  & z_{\ell,3,+} & -z_{\ell,3,-}\\
 &  &  &  & -z_{\ell,3,-} & z_{\ell,3,+}\\
 &  &  &  &  &  & z_{\ell,4,+} & -z_{\ell,4,-}\\
 &  &  &  &  &  & -z_{\ell,4,-} & z_{\ell,4,+}
\end{pmatrix},\label{eq:Vtilde_z}
\end{align}
\begin{equation}
\hat{\mat{\mathcal{V}}}_{\ell}^{\updownarrow}=\begin{pmatrix}t_{\ell,4,+} & \hphantom{-z_{\ell,1,-}} & \hphantom{-z_{\ell,1,-}} & \hphantom{-z_{\ell,1,-}} & \hphantom{-z_{\ell,1,-}} & \hphantom{-z_{\ell,1,-}} & \hphantom{-z_{\ell,1,-}} & -t_{\ell,4,-}\\
\hphantom{-z_{\ell,1,-}} & t_{\ell,1,+} & t_{\ell,1,-} &  &  &  &  & \hphantom{-z_{\ell,1,-}}\\
 & t_{\ell,1,-} & t_{\ell,1,+}\\
 &  &  & t_{\ell,2,+} & t_{\ell,2,-}\\
 &  &  & t_{\ell,2,-} & t_{\ell,2,+}\\
 &  &  &  &  & t_{\ell,3,+} & t_{\ell,3,-}\\
 &  &  &  &  & t_{\ell,3,-} & t_{\ell,3,+}\\
-t_{\ell,4,-} &  &  &  &  &  &  & t_{\ell,4,+}
\end{pmatrix}.\label{eq:Vtilde_t}
\end{equation}
\eS{}\endgroup{}We observe the intuitive picture that alternating
applications $|\hat{\Psi}\rangle\mapsfrom\hat{\mat{\mathcal{V}}}_{\ell}^{\updownarrow}|\hat{\Psi}\rangle$
and $|\hat{\Psi}\rangle\mapsfrom\hat{\mat{\mathcal{V}}}_{\ell}^{\leftrightarrow}|\hat{\Psi}\rangle$
on the state vector $|\hat{\Psi}\rangle$ lead to a repetitive mixing
of its components $|\hat{\Psi}\rangle_{m}$ with left and right neighbor
entries $|\hat{\Psi}\rangle_{m\pm1}$. We now focus on the case of
open BCs in both directions and homogeneous anisotropic couplings.

\section{Open boundary conditions and symmetry \label{sec:Open-boundary-conditions}}

For homogeneous anisotropic couplings $z_{\ell<L,m}=z$, $z_{L,m}=1$,
$t_{\ell,m<M}=t$ and open BCs $t_{\ell,M}=1$ also in vertical direction
we define the symmetric $2\times2$ block transfer matrix
\begin{equation}
\TM 2\Def\begin{bmatrix}\TM + & \TM -\\
\TM - & \TM +
\end{bmatrix}\Def\mat S_{2}\,\mat{\mathcal{V}}_{\L}^{1/2}\mat{\mathcal{V}}_{\M}\mat{\mathcal{V}}_{\L}^{1/2}\,\mat S_{2},\label{eq:T_2}
\end{equation}
where we employed a unitary reversal of the second row and column
with 
\begin{equation}
\mat S_{2}\Def\begin{bmatrix}\mat 1 & \mat 0\\
\mat 0 & \mat S
\end{bmatrix},\quad\mat S\Def\setlength{\arraycolsep}{8pt}\left(\begin{array}{ccc}
 &  & 1\\
 & \iddots\\
1
\end{array}\right),\label{eq:S}
\end{equation}
in order to achieve the highly symmetric structure of $\TM 2$. Below
it will become clear why we denote the two different blocks $\TM{\pm}$.
In terms of $\TM 2$ the partition function (\ref{eq:Z_Det}) becomes
\bS
\begin{equation}
\Det\Def z^{-M}\Det^{\dagger}=\det\langle\mat e_{2}|\TM 2^{L}|\mat e_{2}\rangle,\label{eq:Det(T2)}
\end{equation}
with modified boundary state
\begin{equation}
|\mat e_{2}\rangle\Def\frac{1}{\sqrt{z}}\mat S_{2}\mat{\mathcal{V}}_{\L}^{-1/2}|\mat e\rangle=\frac{1}{\sqrt{2}}|\mat 1\,\mat S\rangle.\label{eq:vec_e}
\end{equation}
\eS{}Note that we have moved an extra factor $z^{M}$ into $C_{2}\Def z^{M}C_{2}^{\dagger}$
to get $|\mat e_{2}\rangle$ independent of $z$.

The two symmetric $M\times M$ blocks of $\TM 2$ are \bS\label{eq:TpTm-abcd}
\begin{equation}
\setlength{\arraycolsep}{8pt}\TM +=\begin{pmatrix}a_{0}^{+} & c\\
c & a & \ddots\\
 & \ddots & \ddots & \ddots\\
 &  & \ddots & a & c\\
 &  &  & c & a_{0}^{-}
\end{pmatrix},\quad\TM -=\begin{pmatrix} &  &  & d^{-} & b_{0}\\
 &  & \iddots & b & d^{+}\\
 & \iddots & \iddots & \iddots\\
d^{-} & b & \iddots\\
b_{\text{0}} & d^{+}
\end{pmatrix},\label{eq:TpTm}
\end{equation}
with matrix elements (cf. (\ref{eq:apm}))
\begin{alignat}{2}
a & =t_{+}z_{+} & b & =-t_{+}z_{-}\nonumber \\
a_{0}^{\pm} & =t_{+}z_{+}+\tfrac{1}{2}(1-t_{+})(z_{+}\pm1) & \qquad b_{0} & =-\tfrac{1}{2}(1+t_{+})z_{-}\label{eq:abcd}\\
c & =-\tfrac{1}{2}t_{-}z_{-} & d^{\pm} & =\pm\tfrac{1}{2}t_{-}(1\pm z_{+}).\nonumber 
\end{alignat}
\eS{}Note that a matrix like $\TM 2$, with X-shaped structure, is
sometimes called a ``cruciform matrix'' and also occurs in the dimer
problem with open BCs \cite{Fisher61}. However, here the components
are tridiagonal and slightly more complicated.

We now turn to the eigensystem $\TM 2\vec{X}_{\lambda}=\lambda\vec{X}_{\lambda}$
of $\TM 2$. Due to the inversion symmetry 
\begin{equation}
\TM 2^{-1}=\begin{bmatrix}\TM + & -\TM -\\
-\TM - & \TM +
\end{bmatrix}\label{eq:T_2inv}
\end{equation}
the $2M$ eigenvalues $\lambda$ occur in pairs $\lambda,\lambda^{-1}$,
and the unitary matrix of normalized eigenvectors $(\mat X)_{\lambda,m}\Def(\vec{X}_{\lambda})_{m}$
can be written as the direct product 
\begin{equation}
\mat X=\mat r_{\frac{\pi}{4}}\otimes\mat x,\label{eq:Xtox}
\end{equation}
with rotation matrix $\mat r_{\theta}$ from (\ref{eq:Rtheta}), provided
that we sort the eigenvalues $\lambda$ of $\TM 2$ in proper order
$\{\lambda_{1},\ldots,\lambda_{M},\lambda_{1}^{-1},\ldots,\lambda_{M}^{-1}\}$,
see below for details on the ordering. Using the $M\times M$ matrix
$\mat x$ together with the corresponding diagonal matrix of eigenvalues,
\begin{equation}
\mat{\Lambda}\Def\diag(\lambda_{1},\ldots,\lambda_{M}),\label{eq:diag_lambda}
\end{equation}
we can define a $M\times M$ transfer matrix 
\begin{equation}
\TM{}\Def\T{\mat x}\mat{\Lambda}\mat x\label{eq:T}
\end{equation}
such that (\ref{eq:T_2}) and (\ref{eq:Xtox}) give
\begin{equation}
\TM{\pm}=\tfrac{1}{2}\left(\TM{}\pm\TM{}^{-1}\right)\quad\Leftrightarrow\quad\TM{}^{\pm1}=\TM +\pm\TM -.\label{eq:T_pm}
\end{equation}
Remarkably, we find $\det\mat{\Lambda}=\det\TM{}=t$. Note that the
$\pm$ notation is as defined in (\ref{eq:apm}). 

We can interpret the steps above as a block diagonalization of $\TM 2$
through a rotation with $\mat R_{\theta}$ from (\ref{eq:Rtheta})
according to 
\begin{equation}
\mat R_{\frac{\pi}{4}}\TM 2\,\T{\mat R}_{\frac{\pi}{4}}=\begin{bmatrix}\TM{} & 0\\
0 & \TM{}^{-1}
\end{bmatrix}.\label{eq:block_diagonal}
\end{equation}
Nonetheless, we first proceed with the simpler tridiagonal matrix
$\TM +$ from (\ref{eq:TpTm}). The eigenvalues of $\TM{\pm}$ fulfill
$\TM{\pm}\vec{x}_{\lambda}=\lambda_{\pm}\vec{x}_{\lambda}$, and we
can analyze the eigensystem of $\TM +$ instead of $\TM 2$ or $\TM{}$,
which is much easier. The eigenvalues $\lambda$ and $\lambda_{\pm}$
are directly related to the Onsager-$\gamma$ via
\begin{equation}
\lambda=\ee^{\gamma},\quad\lambda_{+}=\cosh\gamma,\quad\lambda_{-}=\sinh\gamma.\label{eq:gamma}
\end{equation}

\section{Eigenvalues of $\protect\TM{}$ and the angle $\varphi$}

The characteristic polynomial of the matrix $\TM +$, 
\begin{equation}
P_{M}(\lambda_{+})\Def\det(\TM +-\lambda_{+}\mat 1),\label{eq:CP_def}
\end{equation}
is derived from (\ref{eq:TpTm-abcd}) using the well known recursion
formula for tridiagonal matrices (see, e.g., \cite{Molinari08}),
\begin{align}
P_{M}(\lambda_{+}) & =\langle\begin{array}{cc}
a_{0}^{-}-\lambda_{+}, & c\end{array}|\begin{pmatrix}a-\lambda_{+} & c\\
-c & 0
\end{pmatrix}^{M-2}|\begin{array}{cc}
a_{0}^{+}-\lambda_{+}, & -c\end{array}\rangle\label{eq:CP_def-2}\\
 & =\left(\frac{t_{-}z_{-}}{2}\right)^{M}\langle\begin{array}{cc}
1, & -t^{*}z^{*}\end{array}|\,\mat Q^{M}\,|\begin{array}{cc}
1, & t^{*}/z^{*}\end{array}\rangle,
\end{align}
with 
\begin{equation}
\mat Q=\begin{pmatrix}2\frac{t_{+}z_{+}-\lambda_{+}}{t_{-}z_{-}} & -1\\
1 & 0
\end{pmatrix}.\label{eq:Q}
\end{equation}
The eigenvalues of $\mat Q$,
\begin{equation}
q^{\pm}=\frac{t_{+}z_{+}-\lambda_{+}}{t_{-}z_{-}}\mp\frac{\sqrt{(t_{+}z_{+}-\lambda_{+})^{2}-t_{-}^{2}z_{-}^{2}}}{t_{-}z_{-}}\label{eq:qpm}
\end{equation}
have modulus one and can be written as $q^{\pm}=\ee^{\pm\ii\varphi}$,
if we define the angle $\varphi$ such that
\begin{equation}
\cos\varphi=\frac{t_{+}z_{+}-\lambda_{+}}{t_{-}z_{-}},\quad\sin\varphi=\ii\frac{\sqrt{tz-\lambda}\sqrt{1-tz\lambda}\sqrt{z-t\lambda}\sqrt{t-z\lambda}}{2tz\lambda t_{-}z_{-}}.\label{eq:phidef}
\end{equation}
Note that the factorization of the square root determines the sign
of $\sin\varphi$. Then,
\begin{equation}
\mat Q^{n}=\begin{pmatrix}2\cos\varphi & -1\\
1 & 0
\end{pmatrix}^{n}=\frac{1}{\sin\varphi}\begin{pmatrix}\sin([n+1]\varphi) & -\sin(n\varphi)\\
\sin(n\varphi) & -\sin([n-1]\varphi)
\end{pmatrix},\label{eq:Qn}
\end{equation}
and the characteristic polynomial, now in terms of $\varphi$, simplifies
to 
\begin{equation}
P_{M}(\varphi)=\cos(M\varphi)+\left(t_{+}\cos\varphi-t_{-}\frac{z_{+}}{z_{-}}\right)\frac{\sin(M\varphi)}{\sin\varphi}\label{eq:CP(phi)}
\end{equation}
up to an irrelevant factor $2/(t_{+}+1)(t_{-}z_{-}/2)^{M}$. $P_{M}(\varphi)$
can be written in terms of Chebyshev polynomials of the first and
second kind, $T_{M}(\cos\varphi)=\cos(M\varphi)$ and $U_{M-1}(\cos\varphi)=\sin(M\varphi)/\sin\varphi$,
and is therefore a polynomial of degree $M$ in $\cos\varphi$.

\begin{figure}
\begin{centering}
\includegraphics[width=0.8\columnwidth]{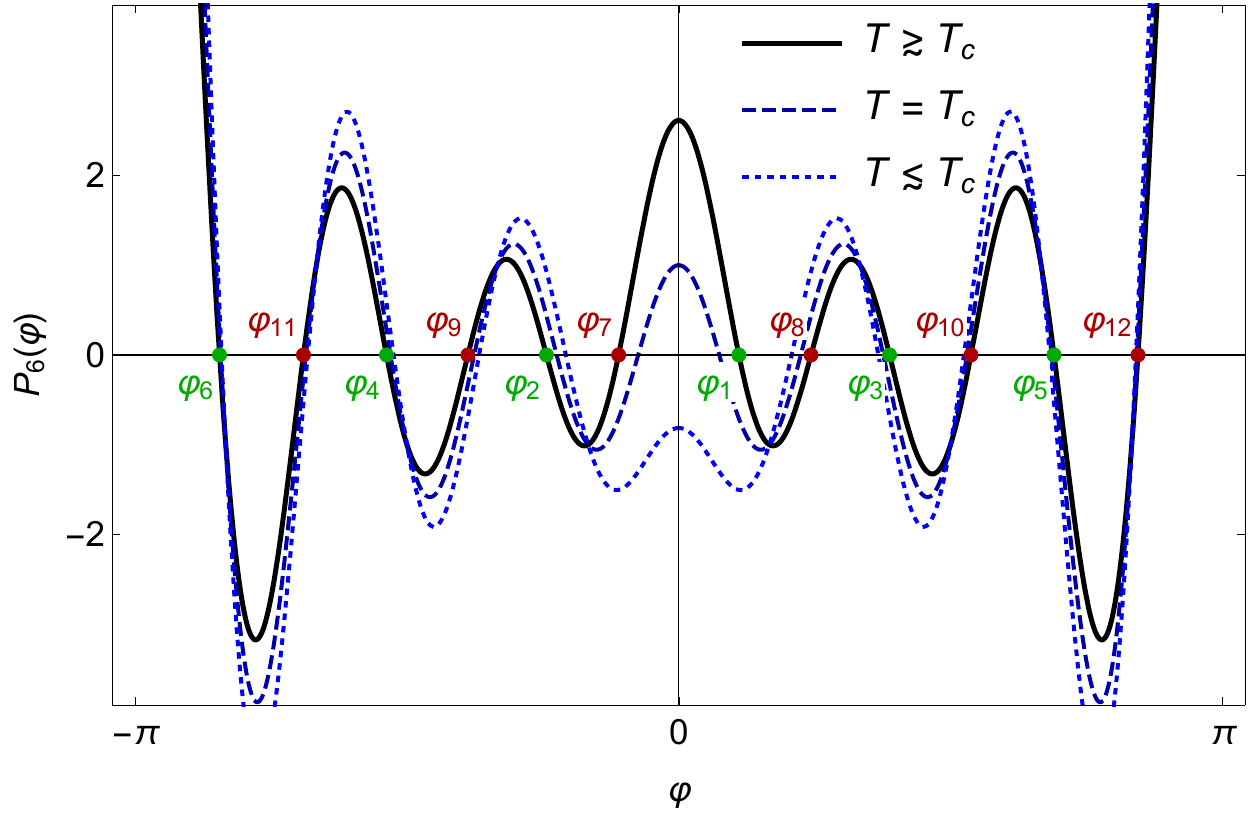}
\par\end{centering}
\caption{Characteristic polynomial $P_{M}(\varphi)$, equation~(\ref{eq:CP(phi)}),
for $M=6$ above, at, and below $\protect\Tc$. The eigenvalues are
ordered as shown (see text).\label{fig:charpoly}}
\end{figure}
Using the characteristic polynomial $P_{M}(\varphi)$ we can come
back to the arrangement of the eigenvalues $\lambda$ of $\TM 2$
and $\TM{}$. It turns out that it is beneficial to sort the $2M$
eigenvalues $\lambda$ of $\TM 2$ by the value of $\varphi$, first
selecting the zeroes of $P_{M}(\varphi)$ with negative slope ordered
by $|\varphi|$ (green points in figure~\ref{fig:charpoly}), and
then selecting the zeroes of $P_{M}(\varphi)$ with positive slope
ordered by $|\varphi|$ (red points in figure~\ref{fig:charpoly}).
Slightly below $\Tc$ the two zeroes $\varphi_{1}$ and $\varphi_{M+1}$
are zero and become complex below \cite{Hucht16b}. However, the corresponding
values $\lambda_{1}$ and $\lambda_{M+1}$ are always real and define
the correct order.

The arrangement is compatible with (\ref{eq:Xtox}) and leads to the
following identities: from equation~(\ref{eq:phidef}), we derive
the identities \bS\label{eq:id1}
\begin{align}
\sin\frac{\varphi}{2} & =-\frac{\sqrt{z-t\lambda}\sqrt{t-z\lambda}}{2\sqrt{tz\lambda}\sqrt{t_{-}z_{-}}},\label{eq:id1-sin}\\
\cos\frac{\varphi}{2} & =\frac{\sqrt{\lambda-tz}\sqrt{1-tz\lambda}}{2\sqrt{tz\lambda}\sqrt{t_{-}z_{-}}},\label{eq:id1-cos}\\
\tan\frac{\varphi}{2} & =-\frac{\sqrt{z-t\lambda}\sqrt{t-z\lambda}}{\sqrt{\lambda-tz}\sqrt{1-tz\lambda}}\label{eq:id1-tan}
\end{align}
\eS{}and, using the characteristic polynomial (\ref{eq:CP(phi)}),
\bS\label{eq:idM}
\begin{align}
\sin\frac{M\varphi}{2} & =\pm\frac{\sqrt{z-t\lambda}\sqrt{1-tz\lambda}}{2\sqrt{tz\lambda}\sqrt{t_{-}\lambda_{-}}},\label{eq:idM-sin}\\
\cos\frac{M\varphi}{2} & =\pm\frac{\sqrt{t-z\lambda}\sqrt{\lambda-tz}}{2\sqrt{tz\lambda}\sqrt{t_{-}\lambda_{-}}},\label{eq:idM-cos}\\
\tan\frac{M\varphi}{2} & =\frac{\sqrt{z-t\lambda}\sqrt{1-tz\lambda}}{\sqrt{t-z\lambda}\sqrt{\lambda-tz}}\label{eq:idM-tan}
\end{align}
\eS{}as well as
\begin{equation}
\frac{\sin(M\varphi)}{\sin\varphi}=-\frac{z_{-}}{\lambda_{-}}.\label{eq:idsin}
\end{equation}
These identities will be used in the following to simplify the eigenvectors
of $\mat{\mathcal{T}}$.

\section{Eigenvectors of $\protect\TM{}$}

The common eigenvectors of $\TM{}$, $\TM +$ and $\TM -$ can be
calculated from the recursion matrix (\ref{eq:Qn}), too, and read
\begin{align}
(\mat x)_{\lambda,n}=(\vec{x}_{\lambda})_{n} & \propto\langle\begin{array}{cc}
1, & 0\end{array}|\,\mat Q^{n}\,|\begin{array}{cc}
1, & t^{*}/z^{*}\end{array}\rangle\nonumber \\
 & \propto\frac{\sin([n+1]\varphi)}{(1-t)(1+z)}-\frac{\sin(n\varphi)}{(1+t)(1-z)},\label{eq:evec_Q}
\end{align}
with $n=0,\ldots,M-1$. After proper normalization and an index change
from $n$ to $m=-M+1,-M+3,\cdots,M-1$, running over the odd integers
between $-M$ and $M$, the matrix elements of $\mat x$ are
\begin{equation}
(\mat x)_{\lambda,m}=\frac{\sqrt{4tz}\,t_{-}z_{-}\lambda_{-}}{\sqrt{M\lambda_{-}^{2}+z_{+}\lambda_{+}-t_{+}}\,\sqrt{\lambda_{+}-1}}\left[\frac{\sin([M+1+m]\frac{\varphi}{2})}{(1-t)(1+z)}-\frac{\sin([M-1+m]\frac{\varphi}{2})}{(1+t)(1-z)}\right].\label{eq:evec_def}
\end{equation}
The block-diagonal transfer matrix (\ref{eq:block_diagonal}) enables
us to reduce the problem of calculating the partition function from
$2M\times2M$ matrices to $M\times M$ matrices, and to factorize
the involved determinants. This will be demonstrated in the following
chapter. 

\section{Partition function factorization}

Using the eigensystem defined above and the block diagonal form (\ref{eq:block_diagonal}),
we can write the partition function (\ref{eq:Det(T2)}) as \bS\label{eq:Det(es)}
\begin{align}
\Det & =\det\langle\mat S^{+}\;\mat S^{-}|\begin{bmatrix}\TM{}^{L} & 0\\
0 & \TM{}^{-L}
\end{bmatrix}|\mat S^{+}\;\mat S^{-}\rangle\label{eq:Det(es)_1}\\
 & =\det\big(\mat S^{+}\TM{}^{L}\,\mat S^{+}+\mat S^{-}\TM{}^{-L}\,\mat S^{-}\big),\label{eq:Det(es)_2}
\end{align}
\eS{}with $\mat S^{\pm}\Def\frac{1}{2}\left(\mat 1\pm\mat S\right)$.
At this point we define the $M\times M$ matrix 
\begin{equation}
\mat M\Def\mat x\big(\TM{}^{L/2}\,\mat S^{+}+\TM{}^{-L/2}\,\mat S^{-}\big),\label{eq:M_def}
\end{equation}
which completes the square in (\ref{eq:Det(es)_2}), as 
\begin{align}
\T{\mat M}\mat M & =\big(\mat S^{+}\TM{}^{L/2}+\mat S^{-}\TM{}^{-L/2}\,\big)\T{\mat x}\,\mat x\big(\TM{}^{L/2}\,\mat S^{+}+\TM{}^{-L/2}\,\mat S^{-}\big)\nonumber \\
 & =\mat S^{+}\TM{}^{L}\,\mat S^{+}+\mat S^{+}\mat S^{-}+\mat S^{-}\mat S^{+}+\mat S^{-}\TM{}^{-L}\,\mat S^{-}\nonumber \\
 & =\mat S^{+}\TM{}^{L}\,\mat S^{+}+\mat S^{-}\TM{}^{-L}\,\mat S^{-}\label{eq:MM-is-square}
\end{align}
and the mixed terms in the expansion vanish, $\mat S^{+}\mat S^{-}=\mat S^{-}\mat S^{+}=\frac{1}{4}\left(\mat 1-\mat S^{2}\right)=\mat 0$.
With $\mat x\TM{}^{\pm L/2}=\mat{\Lambda}^{\pm L/2}\mat x$ from (\ref{eq:T})
the matrix elements of $\mat M$ are 
\begin{equation}
(\mat M)_{\lambda,m}=\tfrac{1}{2}(\lambda^{L/2}+\lambda^{-L/2}\,)(\mat x)_{\lambda,m}+\tfrac{1}{2}(\lambda^{L/2}-\lambda^{-L/2}\,)(\mat x)_{\lambda,-m},\label{eq:M_1}
\end{equation}
and the partition function (\ref{eq:Det(es)}) becomes
\begin{equation}
\Det=\det\left(\T{\mat M}\mat M\right)={\textstyle \det^{2}\mat M},\label{eq:Det(M)}
\end{equation}
i.\,e.~$Z\propto\det\mat M$.

We now insert the definition of $\mat x$ from (\ref{eq:evec_def})
and pull out common $m$-independent factors, primarily the normalization
constants, which we can move into a diagonal matrix $\mat D$ according
to 
\begin{equation}
\T{\mat M}\mat M\Def\T{\mat W}\mat D\mat W.\label{eq:WDW_def}
\end{equation}
We first choose the decomposition \bS
\begin{align}
(\mat W^{\ddagger})_{\lambda,m} & \Def\frac{1}{2}\sum_{\pm}(\lambda^{L/2}\pm\lambda^{-L/2})\left(\frac{\sin([M+1\pm m]\frac{\varphi}{2})}{(1-t)(1+z)}-\frac{\sin([M-1\pm m]\frac{\varphi}{2})}{(1+t)(1-z)}\right),\label{eq:W'_def}\\
(\mat D^{\ddagger})_{\lambda,\lambda} & \Def\frac{8tz\lambda\,(t_{-}z_{-}\lambda_{-})^{2}}{(M\lambda_{-}^{2}+z_{+}\lambda_{+}-t_{+})\,(1-\lambda)^{2}},\label{eq:D'_def}
\end{align}
\eS{}and sort $(\mat W^{\ddagger})_{\lambda,m}$ by terms in $\lambda^{\pm L/2}$
to get, after some trigonometry,
\begin{align}
(\mat W^{\ddagger})_{\lambda,m}=\frac{\sin\varphi}{4tt_{-}zz_{-}} & \left[\lambda^{L/2}\left((t-z)\frac{\sin\frac{M\varphi}{2}}{\sin\frac{\varphi}{2}}-(tz-1)\frac{\cos\frac{M\varphi}{2}}{\cos\frac{\varphi}{2}}\right)\cos\frac{m\varphi}{2}\right.\nonumber \\
 & {}+\left.\lambda^{-L/2}\left((t-z)\frac{\cos\frac{M\varphi}{2}}{\sin\frac{\varphi}{2}}+(tz-1)\frac{\sin\frac{M\varphi}{2}}{\cos\frac{\varphi}{2}}\right)\sin\frac{m\varphi}{2}\right].\label{eq:W'_1}
\end{align}
Pulling out some factors and rearranging terms we get
\begin{align}
(\mat W^{\ddagger})_{\lambda,m}=\frac{\sin\varphi\cos\frac{M\varphi}{2}}{4tt_{-}zz_{-}} & \left[\lambda^{L/2}\left((t-z)\frac{\tan\frac{M\varphi}{2}}{\tan\frac{\varphi}{2}}-(tz-1)\right)\frac{\cos\frac{m\varphi}{2}}{\cos\frac{\varphi}{2}}+{}\right.\nonumber \\
 & {}+\left.\lambda^{-L/2}\left((t-z)+(tz-1)\frac{\tan\frac{M\varphi}{2}}{\cot\frac{\varphi}{2}}\right)\frac{\sin\frac{m\varphi}{2}}{\sin\frac{\varphi}{2}}\right].\label{eq:W'_2}
\end{align}
Further simplifications occur if we use the identities from (\ref{eq:id1})
and (\ref{eq:idM}), especially 
\begin{equation}
\frac{\tan\frac{M\varphi}{2}}{\cot\frac{\varphi}{2}}=\frac{z-t\lambda}{tz-\lambda},\qquad\frac{\tan\frac{M\varphi}{2}}{\tan\frac{\varphi}{2}}=\frac{tz\lambda-1}{t-z\lambda}.\label{eq:id_tan}
\end{equation}
Shifting again $m$-independent factors from $\mat W^{\ddagger}$
to $\mat D^{\ddagger}$, the result can be simplified to \bS
\begin{align}
(\mat W^{\dagger})_{\lambda,m} & \Def\frac{1}{\sqrt{t_{-}z_{-}}}\left[\lambda^{L/2}(tz-\lambda)\frac{\cos\frac{m\varphi}{2}}{\cos\frac{\varphi}{2}}-\lambda^{-L/2}(tz^{-1}-\lambda)\frac{\sin\frac{m\varphi}{2}}{\sin\frac{\varphi}{2}}\right]\label{eq:W}\\
(\mat D)_{\lambda,\lambda} & \Def\frac{\lambda_{-}}{2z_{-}}\frac{(t_{+}z_{+}-\lambda_{+})^{2}-t_{-}^{2}z_{-}^{2}}{M\lambda_{-}^{2}+z_{+}\lambda_{+}-t_{+}}\,\frac{1}{(tz-\lambda)(tz^{-1}-\lambda)},\label{eq:D}
\end{align}
\eS{}and equation (\ref{eq:Det(M)}) becomes 
\begin{equation}
\Det={\det}^{2}\mat W^{\dagger}\;\prod_{\lambda}(\mat D)_{\lambda\lambda}.\label{eq:Det(W)}
\end{equation}
The remaining challenge is the calculation of $\det\mat W^{\dagger}$,
which will be further simplified in the following.

\section{The Vandermonde determinant}

We now utilize the observation that the matrix $\mat W^{\dagger}$
is a Vandermonde matrix, and that its determinant is invariant under
basis transformations between complete polynomial bases. Hence we
can transform $\mat W^{\dagger}$ from the trigonometric basis to
the simpler power basis. We identify the leading term in both $\cos\frac{m\varphi}{2}/\cos\frac{\varphi}{2}$
and $\sin\frac{m\varphi}{2}/\sin\frac{\varphi}{2}$ to be\footnote{``$\simeq$'' denoted ``asymptotically equal''}
\begin{equation}
\frac{\cos\frac{m\varphi}{2}}{\cos\frac{\varphi}{2}}\simeq\left(2\cos\frac{\varphi}{2}\right)^{|m|-1},\qquad\frac{\sin\frac{m\varphi}{2}}{\sin\frac{\varphi}{2}}\simeq\frac{m}{|m|}\left(2\cos\frac{\varphi}{2}\right)^{|m|-1}\label{eq:cos_sin_leading}
\end{equation}
and rewrite the result using (\ref{eq:id1-cos}), as $2n\Def|m|-1$
is an even integer, to
\begin{equation}
\left(2\cos\frac{\varphi}{2}\right)^{2n}=\left[\frac{(\lambda-tz)(1-tz\lambda)}{tz\lambda t_{-}z_{-}}\right]^{n}\simeq\left(\frac{-2}{t_{-}z_{-}}\right)^{n}\lambda_{+}^{n}.\label{eq:cos_leading}
\end{equation}
The determinant becomes 
\begin{equation}
\det\mat W^{\dagger}=\left(\frac{2}{t_{-}z_{-}}\right)^{M^{2}/2}\det\mat W\label{eq:detW-dethatW}
\end{equation}
 with \begingroup\renewcommand*{\arraystretch}{1.1}
\begin{equation}
\mat W=\left(\begin{array}{cccc|cccc}
g_{1}c_{1}^{M/2-1} & \cdots & g_{1}c_{1} & g_{1} & f_{1} & f_{1}c_{1} & \cdots & f_{1}c_{1}^{M/2-1}\\
g_{2}c_{2}^{M/2-1} & \cdots & g_{2}c_{2} & g_{2} & f_{2} & f_{2}c_{2} & \cdots & f_{2}c_{2}^{M/2-1}\\
\vdots &  & \vdots & \vdots & \vdots & \vdots &  & \vdots\\
g_{M}c_{M}^{M/2-1} & \cdots & g_{M}c_{M} & g_{M} & f_{M} & f_{M}c_{M} & \cdots & f_{M}c_{M}^{M/2-1}
\end{array}\right),\label{eq:hatW}
\end{equation}
\endgroup where we introduced the abbreviations
\begin{equation}
c_{\mu}\Def\lambda_{\mu,+},\quad g_{\mu}\Def-\lambda_{\mu}^{L/2}(tz-\lambda_{\mu}),\quad f_{\mu}\Def\lambda_{\mu}^{-L/2}(tz^{-1}-\lambda_{\mu}).\label{eq:gf_def}
\end{equation}
Using a block Laplace expansion along the vertical line in (\ref{eq:hatW}),
the determinant of $\mat W$ can be written as alternating sum over
all possible $\Mh\times\Mh$ $g$-minors $\det\mat W_{\S,\{1,\ldots,\Mh\}}$,
times the corresponding $f$-minors $\det\mat W_{\Sc,\{\Mh+1,\ldots,M\}}$,
\begin{equation}
\det\mat W=\pm\sum_{\S}\sign(\S,\Sc)\underbrace{\prod_{\mu\in\S}g_{\mu}\:\prod_{\mu<\nu\in\S}\negmedspace(c_{\mu}-c_{\nu})}_{\det\mat W_{\S,\{1,\ldots,\Mh\}}}\:\underbrace{\prod_{\mu\in\Sc}f_{\mu}\:\prod_{\mu<\nu\in\Sc}\negmedspace(c_{\mu}-c_{\nu})}_{\det\mat W_{\Sc,\{\Mh+1,\ldots,M\}}},\label{eq:hatW_expansion}
\end{equation}
where $\S$ denotes one of the $\binom{M}{M/2}$ possible subsets
of $\Mh$ choices of the index set $\{1,\ldots,M\}$, and $\Sc$ its
complement. Both minors are simple Vandermonde determinants, and the
irrelevant overall sign depends on the ordering within the sets. 

In the following, we further reduce the matrix size from $M\times M$
to $\Mh\times\Mh$ by Vandermonde-type row elimination. While for
simple Vandermonde determinants this procedure leads a complete factorization,
in our case we can only eliminate $\Mh$ rows, which we nevertheless
can choose arbitrary. We now denote the chosen set of eliminated rows
and its complement by $\S$ and $\Sc$ and find ($\mat A_{\S}\Def\mat A_{\S,\S}$)
\begin{equation}
\det\mat W=\pm d_{\S,\Sc}\det\left(\mat G_{\S}\mat T_{\S,\Sc}\mat F_{\Sc}-\mat F_{\S}\mat T_{\S,\Sc}\mat G_{\Sc}\right),\label{eq:hatW_1}
\end{equation}
with the $M\times M$ matrices \bS 
\begin{equation}
(\mat G)_{\mu\mu}\Def g_{\mu},\quad(\mat F)_{\mu\mu}\Def f_{\mu},\quad(\mat T)_{\mu\nu}\Def\frac{1}{c_{\mu}-c_{\nu}},\label{eq:GFT_def}
\end{equation}
(we can set $(\mat T)_{\mu\mu}\Def0$, as $\mu\neq\nu$), and with
the double product
\begin{equation}
d_{\S,\Sc}\Def\prod_{\mu\in\S}\prod_{\nu\in\Sc}(c_{\mu}-c_{\nu}).\label{eq:d_SSc}
\end{equation}
\eS{}As $\mat T_{\S,\Sc}$ is a Cauchy matrix, both $\mat G_{\S}\mat T_{\S,\Sc}\mat F_{\Sc}$
and $\mat F_{\S}\mat T_{\S,\Sc}\mat G_{\Sc}$ are Cauchy-like matrices.
An example with $M=6$ and $\S=\{1,3,5\}$, such that $\Sc=\{2,4,6\}$,
reads
\begin{equation}
\mat G_{\S}=\left(\begin{array}{ccc}
g_{1}\\
 & g_{3}\\
 &  & g_{5}
\end{array}\right),\quad\mat F_{\Sc}=\left(\begin{array}{ccc}
f_{2}\\
 & f_{4}\\
 &  & f_{6}
\end{array}\right),\quad\mat T_{\S,\Sc}=\left(\begin{array}{ccc}
\frac{1}{c_{1}-c_{2}} & \frac{1}{c_{1}-c_{4}} & \frac{1}{c_{1}-c_{6}}\\
\frac{1}{c_{3}-c_{2}} & \frac{1}{c_{3}-c_{4}} & \frac{1}{c_{3}-c_{6}}\\
\frac{1}{c_{5}-c_{2}} & \frac{1}{c_{5}-c_{4}} & \frac{1}{c_{5}-c_{6}}
\end{array}\right).\label{eq:GFT-example}
\end{equation}
The choice of $\S$ has influence on the magnitude of the two terms
in (\ref{eq:hatW_1}) and has a physical interpretation: if we choose
$\S=\So\Def\{1,3,\ldots,M-1\}$ odd integers, both $\mat G_{\S}$
and $\mat F_{\Sc}$ contain only dominant (for large $L$) eigenvalues
$\lambda_{\mu}>1$, while the subdominant ones $\lambda_{\mu}<1$
enter $\mat G_{\Sc}$ and $\mat F_{\S}$. Therefore, the term $\mat G_{\S}\mat T_{\S,\Sc}\mat F_{\Sc}$
in (\ref{eq:hatW_1}) gives the leading contribution for large $L$,
and the second one $\mat F_{\S}\mat T_{\S,\Sc}\mat G_{\Sc}$ the finite-$L$
corrections. The oscillating behavior 
\begin{equation}
\sign\log\lambda_{\mu}=\sign\gamma_{\mu}=\sign\varphi_{\mu}=(-1)^{\mu-1},\quad\mu=1,\ldots,M,\label{eq:oscillating}
\end{equation}
is dictated by the ordering of the zeroes of $P_{M}(\varphi)$, equation~(\ref{eq:CP(phi)}),
as described above. 

Consequently, we factor out the leading first term of the determinant
in (\ref{eq:hatW_1}),
\begin{equation}
\det\mat W=\pm d_{\S,\Sc}\det\big(\mat G_{\S}\mat T_{\S,\Sc}\mat F_{\Sc}\big)\det\big(\mat 1-\mat F_{\Sc}^{-1}\mat T_{\S,\Sc}^{-1}\mat G_{\S}^{-1}\mat F_{\S}\mat T_{\S,\Sc}\mat G_{\Sc}\big),\label{eq:hatW_2}
\end{equation}
and express the inverse $\mat T_{\S,\Sc}^{-1}$ through the diagonal
matrix
\begin{equation}
(\mat P)_{\mu\mu}\Def p_{\mu}\Def\prod_{\nu=1}^{M}{\vphantom{\prod}}'(c_{\mu}-c_{\nu})^{-\sigma_{\mu}\sigma_{\nu}},\label{eq:P_mumu_def}
\end{equation}
which fulfills
\begin{equation}
\mat P_{\Sc}\mat T_{\Sc,\S}\mat P_{\S}\mat T_{\S,\Sc}=\mat 1.\label{eq:PTPT}
\end{equation}
Here, $\prod'$ denotes the regularized product, with zero and infinite
factors removed, and we have defined the parity of $\mu$
\begin{equation}
\sigma_{\mu}\Def\left\{ \begin{array}{cc}
+1 & \quad\text{if }\mu\in\S\hphantom{.}\\
-1 & \quad\text{if }\mu\in\Sc.
\end{array}\right.\label{eq:sigma_mu}
\end{equation}
We now introduce the diagonal matrix 
\begin{equation}
(\mat V)_{\mu\mu}\Def v_{\mu}\Def-p_{\mu}\lambda_{\mu}^{L}\left(\frac{g_{\mu}}{f_{\mu}}\right)^{\sigma_{\mu}}=p_{\mu}\frac{tz^{-\sigma_{\mu}}-\lambda_{\mu}}{tz^{\sigma_{\mu}}-\lambda_{\mu}}\label{eq:V-def}
\end{equation}
and define, with $\mat{\Lambda}$ from (\ref{eq:diag_lambda}), for
the specific set of dominant odd indices $\So$ as well as the complementary
set of even indices $\Soc\Def\bar{\So}$ the residual matrix
\begin{equation}
\mat Y\Def-\mat{\Lambda}_{\Soc}^{L}\mat V_{\Soc}\mat T_{\Soc,\So}\mat{\Lambda}_{\So}^{-L}\mat V_{\So}\mat T_{\So,\Soc}\label{eq:Y}
\end{equation}
to find
\begin{equation}
\det\mat W=\pm d_{\So,\Soc}\det\mat T_{\So,\Soc}\det\mat G_{\So}\det\mat F_{\Soc}\det(\mat 1+\mat Y).\label{eq:hatW_3}
\end{equation}
Remember that the matrices with one index are diagonal. The determinant
of the Cauchy matrix $\mat T_{\S,\Sc}$ reads 
\begin{equation}
\det\mat T_{\S,\Sc}=\pm\frac{q_{\S}q_{\Sc}}{d_{\S,\Sc}},\label{eq:detT_SSc}
\end{equation}
with 
\begin{equation}
q_{\S}\Def\prod_{\mu<\nu\in\S}\negmedspace(c_{\mu}-c_{\nu}),\label{eq:q_S_def}
\end{equation}
leading to the final form
\begin{equation}
\det\mat W=|q_{\So}q_{\Soc}|\det\mat G_{\So}\det\mat F_{\Soc}\det(\mat 1+\mat Y).\label{eq:dethatW}
\end{equation}

\section{Resulting partition function}

Introducing the strip residual partition function
\begin{equation}
\Zsres\Def\det(\mat 1+\mat Y)\label{eq:Zsres}
\end{equation}
for the remaining determinant, and inserting explicit values for $\det\mat G_{\So}$,
$\det\mat F_{\Soc}$ and $\det\mat D$, we arrive at the final result
\bS
\begin{equation}
Z=\Bigg[C_{3}\,d_{\So,\Soc}^{2}\,\prod_{\mu=1}^{M}\frac{(t_{+}z_{+}-\lambda_{\mu,+})^{2}-t_{-}^{2}z_{-}^{2}}{M\lambda_{\mu,-}^{2}+z_{+}\lambda_{\mu,+}-t_{+}}\,\frac{\sigma_{\mu}\lambda_{\mu,-}}{v_{\mu}}\,\lambda_{\mu}^{\sigma_{\mu}L}\Bigg]^{1/2}\:\Zsres\label{eq:Z-final}
\end{equation}
for the partition function, with parity $\sigma_{\mu}=(-1)^{\mu-1}$,
$d_{\So,\Soc}$ from (\ref{eq:d_SSc}), and the constant
\begin{equation}
C_{3}\Def z^{M}\left(\frac{2}{z_{-}}\right)^{LM}\left(\frac{2}{t_{-}z_{-}}\right)^{M^{2}/2}.\label{eq:K_3}
\end{equation}
\eS{}We can discuss two limiting cases with respect to the aspect
ratio $\rho$: by definition, the matrix $\mat Y$ only contains subdominant
finite-$L$ contributions, and therefore $\lim_{\rho\to\infty}\mat Y=\mat 0$
and $\lim_{\rho\to\infty}\Zsres=1$. On the other hand, the limit
$\rho\to0$ can also be discussed. As $L$ is a real number in (\ref{eq:Y}),
we can let $L\to0$ and find a Cauchy-type matrix very similar to
one describing the spontaneous magnetization of the superintegrable
chiral Potts model, as discussed by Baxter \cite{Baxter10a}. The
resulting determinant can be factorized and reads 
\begin{equation}
\lim_{\rho\to0}\Zsres=\pm\frac{(2tz_{-})^{M/2}{\displaystyle \prod_{\mu\in\So}\prod_{\nu\in\Soc}(\lambda_{\mu}-\lambda_{\nu})}}{{\displaystyle \prod_{\mu=1}^{M}(tz^{\sigma_{\mu}}-\lambda_{\mu})\prod_{\mu<\nu\in\So}(1-\lambda_{\mu}\lambda_{\nu})\prod_{\mu<\nu\in\Soc}(1-\lambda_{\mu}\lambda_{\nu})}}.\label{eq:Zsres_0}
\end{equation}
To summarize, we find closed product representations for both limit
cases $L/M\to\infty$ and $M/L\to\infty$ with finite $M$. The general
case $0<\rho<\infty$, however, involves the nontrivial determinant
(\ref{eq:Zsres}).

The oscillating order of the eigenvalues introduced in Chapter \ref{sec:Open-boundary-conditions}
was a prerequisite for the simple block diagonalization of the block
transfer matrix $\TM 2$, equation~(\ref{eq:block_diagonal}), and
the subsequent factorization of $Z$. However, now we observe that
this oscillation is reversed by the sets $\So$ and $\Soc$ of odd
and even indices, used in the definition of the residual matrix $\mat Y$.
Therefore, we rewrite the results (\ref{eq:Y}) and (\ref{eq:Z-final})
in terms of the simpler non-oscillating dominant eigenvalues $\dom{\lambda}_{\mu}$.
Using the parity $\sigma_{\mu}=(-1)^{\mu-1}$, we define\footnote{remember that $\varphi_{1}$ becomes imaginary below $\Tc$}
\begin{equation}
\dom{\lambda}_{\mu}\Def\lambda_{\mu}^{\sigma_{\mu}}>1,\quad\dom{\gamma}_{\mu}\Def\sigma_{\mu}\gamma_{\mu}=|\gamma_{\mu}|>0,\quad\dom{\varphi}_{\mu}\Def\sigma_{\mu}\varphi_{\mu}\stackrel{\mu>1}{=}|\varphi_{\mu}|>0,\quad\mu=1,\ldots,M,\label{eq:lam-gam-phi-tilde}
\end{equation}
implying $\dom{\lambda}_{\mu,+}=\lambda_{\mu,+}=\dom c_{\mu}=c_{\mu}$
and $\dom{\lambda}_{\mu,-}=\sigma_{\mu}\lambda_{\mu,-}=|\lambda_{\mu,-}|$,
to get \bS\label{eq:Z-final-tilde-all}
\begin{align}
\dom v_{\mu} & =v_{\mu}=p_{\mu}\frac{tz^{-\sigma_{\mu}}-\dom{\lambda}_{\mu}^{\sigma_{\mu}}}{tz^{\sigma_{\mu}}-\dom{\lambda}_{\mu}^{\sigma_{\mu}}},\label{eq:V-tilde}\\
\dom{\mat Y} & =\mat Y=-\dom{\mat{\Lambda}}_{\Soc}^{-L}\mat V_{\Soc}\mat T_{\Soc,\So}\dom{\mat{\Lambda}}_{\So}^{-L}\mat V_{\So}\mat T_{\So,\Soc},\label{eq:Y-tilde}
\end{align}
leading to the partition function (\ref{eq:Z-final}) in terms of
$\dom{\lambda}_{\mu}$,
\begin{equation}
Z=\Bigg[C_{3}\,d_{\So,\Soc}^{2}\,\prod_{\mu=1}^{M}\frac{(t_{+}z_{+}-\dom{\lambda}_{\mu,+})^{2}-t_{-}^{2}z_{-}^{2}}{M\dom{\lambda}_{\mu,-}^{2}+z_{+}\dom{\lambda}_{\mu,+}-t_{+}}\,\frac{\dom{\lambda}_{\mu,-}}{v_{\mu}}\,\dom{\lambda}_{\mu}^{L}\Bigg]^{1/2}\det(\mat 1+\mat Y).\label{eq:Z-final-tilde}
\end{equation}
\eS{}This is the final result of our analysis for arbitrary temperature
$T$ and finite system size $L$ and $M$. We factorized the partition
function up to the factor $\Zsres$, equation~(\ref{eq:Zsres}),
where the residual matrix $\mat Y$ contains all information about
the finite aspect ratio $\rho$ and will be analyzed in detail in
\cite{Hucht16b}. The first term in (\ref{eq:Z-final-tilde}) is the
infinite strip contribution, which has been analyzed in great detail
by Baxter recently \cite{Baxter16}. 

\section{Free energy contributions}

In this chapter we give a decomposition of the reduced free energy
(in units of $k_{\mathrm{B}}T$) 
\begin{equation}
F(T;L,M)=-\log Z\label{eq:F}
\end{equation}
appropriate for our geometry and method. We first recall that 
\begin{equation}
F(T;L,M)=\Finf(T;L,M)+\Fres(T;L,M),\label{eq:F1}
\end{equation}
with infinite volume contribution $\Finf$, that for our geometry
has the form
\begin{equation}
\Finf(T;L,M)\Def LM\fb(T)+L\fs^{\L}(T)+M\fs^{\M}(T)+\fc(T),\label{eq:F_inf}
\end{equation}
and can be viewed as a regularization term in the limit $L,M\to\infty$.
The bulk free energy per spin $\fb(T)$, surface free energies per
surface spin pair $\fs^{\delta}(T)$, and corner free energy $\fc(T)$
are defined in the thermodynamic limit $L,M\to\infty$ and do not
depend on $L,M$. However, the residual free energy $\Fres$, denoted
$O(\ee^{-\gamma L},\ee^{-\gamma M})$ in equation~(1.1) of \cite{Baxter16},
gives rise to important finite-size effects, most prominently the
Casimir amplitude and the critical Casimir force \cite{Hucht16b}. 

In the limit $L\to\infty$ with fixed $M$, the strip residual partition
function $\Zsres\to1$, as shown in the last chapter. Consequently,
we denote the infinite strip contribution 
\begin{equation}
\Zstrip\Def Z/\Zsres\label{eq:Z_inf^L-def}
\end{equation}
and get a free energy decomposition slightly different from (\ref{eq:F1}),
namely
\begin{equation}
F(T;L,M)=\Fstrip(T;L,M)+\Fsres(T;L,M),\label{eq:F1-1}
\end{equation}
where we can identify the strip residual free energy 
\begin{equation}
\Fsres\Def-\log\det(\mat 1+\mat Y)\label{eq:F_res^L-def}
\end{equation}
as the $L$-dependent term in the difference between the residual
free energy $\Fres$ of the finite rectangular system and the leading
divergent term $\mathcal{O}(L)$ in the limit $L\to\infty$ \cite{HuchtGruenebergSchmidt11},
\begin{equation}
\Fsres(T;L,M)=\Fres(T;L,M)-L\lim_{L\to\infty}L^{-1}\Fres(T;L,M)-F_{\mathrm{s,c}}^{\mathrm{res}}(T;M).\label{eq:F_strip^res}
\end{equation}
Note that the last term $F_{\mathrm{s,c}}^{\mathrm{res}}(T;M)$ drops
out in the $L$-derivative below, for details we refer to \cite{Hucht16b}.
In this notation, Vernier \& Jacobsen \cite{VernierJacobsen12} conjectured
a product representation of the infinite volume contribution $\Zinf\Def\ee^{-\Finf}$
that trivially depends on the system size $L$ and $M$, see also
appendix~\ref{apx:Product-formulas}, while Baxter derived a product
formula for the infinite strip contribution $\Zstrip$ at finite strip
width $M$, and then performed the limit $M\to\infty$ \cite{Baxter16}.
Both results applied only to the ordered phase below $\Tc$. 

Finally we turn to the critical Casimir force. The reduced Casimir
force per area $M$ in $L$ direction reads
\begin{equation}
\FC(T;L,M)\Def-\frac{1}{M}\frac{\partial}{\partial L}\Fres(T;L,M)\label{eq:FC^L}
\end{equation}
and can be decomposed into two parts to find, in analogy to (\ref{eq:F_strip^res}),
the differential contribution \bS
\begin{align}
\FC_{\strip}(T;L,M) & \Def-\frac{1}{M}\frac{\partial}{\partial L}\Fsres(T;L,M)\label{eq:dFC^L_a}\\
 & =\FC(T;L,M)+\frac{1}{M}\lim_{L\to\infty}L^{-1}\Fres(T;L,M).\label{eq:dFC^L_b}
\end{align}
\eS{}This contribution is therefore directly related to the remaining
determinant (\ref{eq:Zsres}). For details on the involved universal
amplitudes and finite-size scaling functions the reader again is referred
to \cite{Hucht16b}.

\section{Effective spin model}

\begin{figure}
\begin{centering}
\includegraphics[scale=0.45]{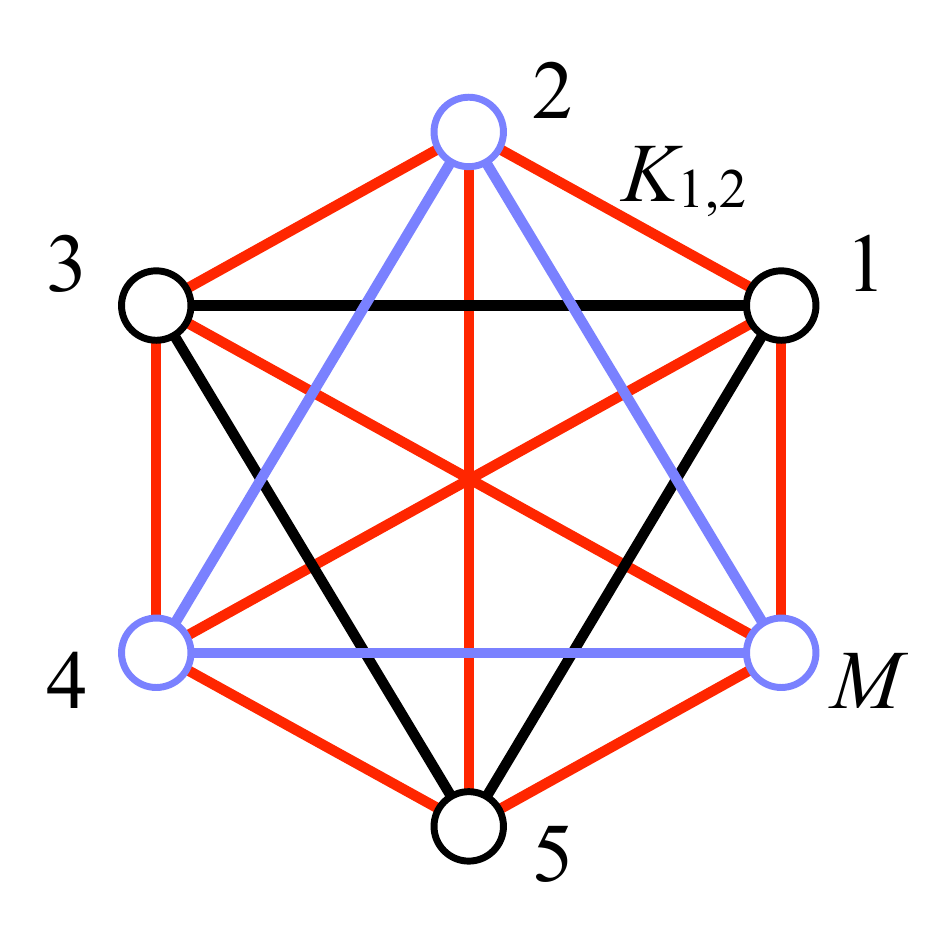}
\par\end{centering}
\caption{Effective spin model for $M=6$. The two sublattices of odd and even
spins are shown as black and light blue circles. The black and light
blue interactions between spins of same parity are ferromagnetic,
while the red couplings are antiferromagnetic. Note that the spatial
arrangement of the spins is arbitrary, as all couplings $K_{\mu\nu}$
are different.\label{fig:Effective-spin-model}}
\end{figure}
In this last chapter we present an exact mapping of the residual determinant
$\Zsres$, equation~(\ref{eq:Zsres}), onto an effective spin model
with $M$ spins and long-range pair interactions. This model might
be a starting point for further investigations of the residual determinant.
The mapping is motivated by the observation that the determinant expansion
of (\ref{eq:Zsres}) is of the form (here we set $L=0$ for simplicity)
\begin{equation}
\Zsres=1+\underbrace{\sum_{\mu\in\So}\sum_{\nu\in\Soc}\frac{v_{\mu}v_{\nu}}{(c_{\mu}-c_{\nu})^{2}}}_{1^{\mathrm{st}}\,\mathrm{order}}+\underbrace{\sum_{\mu\neq\mu'\in\So}\sum_{\nu\neq\nu'\in\Soc}\frac{v_{\mu}v_{\mu'}v_{\nu}v_{\nu'}(c_{\mu}-c_{\mu'})^{2}(c_{\nu}-c_{\nu'})^{2}}{(c_{\mu}-c_{\nu})^{2}(c_{\mu}-c_{\nu'})^{2}(c_{\mu'}-c_{\nu})^{2}(c_{\mu'}-c_{\nu'})^{2}}}_{2^{\mathrm{nd}}\,\mathrm{order}}+\ldots\label{eq:Zsres-expansion}
\end{equation}
and consists of $\binom{M}{M/2}$ positive terms. Hence we identify
these terms with the Boltzmann factors $\ee^{-\mathcal{H}_{\mathrm{eff}}}$
of the $\binom{M}{M/2}$ possible spin configurations of $M$ spins
$s_{\mu}\in\{0,1\}$ under the constraint 
\begin{equation}
\sum_{\mu\in\So}s_{\mu}=\sum_{\nu\in\Soc}s_{\nu}\quad\Leftrightarrow\quad\sum_{\mu=1}^{M}\sigma_{\mu}s_{\mu}=0.\label{eq:constraint-eff-model}
\end{equation}
We interpret $\So$ and $\Soc$ as two sublattices, discriminated
by the parity $\sigma_{\mu}$, equation~(\ref{eq:sigma_mu}), see
figure~\ref{fig:Effective-spin-model}. The effective spin model
then has the Hamiltonian
\begin{equation}
\mathcal{H}_{\mathrm{eff}}=-\sum_{\mu<\nu=1}^{M}K_{\mu\nu}s_{\mu}s_{\nu}+L\sum_{\mu=1}^{M}\dom{\gamma}_{\mu}s_{\mu}+b\Big[\sum_{\mu=1}^{M}\sigma_{\mu}s_{\mu}\Big]^{2},\label{eq:H-eff}
\end{equation}
with interaction constants 
\begin{equation}
K_{\mu\nu}=-\sigma_{\mu}\sigma_{\nu}\log\frac{v_{\mu}v_{\nu}}{(c_{\mu}-c_{\nu})^{2}},\label{eq:K_munu}
\end{equation}
while the positive $\dom{\gamma}_{\mu}$ from (\ref{eq:lam-gam-phi-tilde})
play the role of magnetic moments in a homogeneous magnetic field
of strength $-L$. Both the couplings $K_{\mu\nu}$ as well as the
magnetic moments $\dom{\gamma}_{\mu}$ depend on the temperature of
the underlying Ising model, and the limit $b\to\infty$ enforces the
constraint (\ref{eq:constraint-eff-model}). As $(c_{\mu}-c_{\nu})^{2}>v_{\mu}v_{\nu}$
for all $\mu,\nu$, the couplings $K_{\mu\nu}$ are ferromagnetic
for spins within the same set and antiferromagnetic between different
sets, as shown in figure~\ref{fig:Effective-spin-model}. For $L>0$,
the external magnetic field is antiparallel to the spins and favors
states with small magnetization. Consequently, for magnetic field
$L\to\infty$ all spins are forced to have $s_{\mu}=0$.

With these definitions, the residual determinant (\ref{eq:Zsres})
is equal to the partition function of the Hamiltonian (\ref{eq:H-eff})
in the limit $b\to\infty$,
\begin{equation}
\Zsres=Z_{\mathrm{eff}}\Def\Tr\ee^{-\mathcal{H}_{\mathrm{eff}}},\label{eq:detY-equals-eff-model}
\end{equation}
where the trace effectively runs over the $\binom{M}{M/2}$ spin states
compatible with condition (\ref{eq:constraint-eff-model}), and (\ref{eq:Zsres-expansion})
coincides with the expansion of $Z_{\mathrm{eff}}$ around the high-field
limit $L\to\infty$. In this expansion we start with $s_{\mu}=0$
($Z_{\mathrm{eff}}=1$) and flip one spin in both sublattices to get
the first order term. For two reversed spins in both subsystems we
find the second order term, and so on. 

On the other hand, the zero-field case $L=0$ is described by (\ref{eq:Zsres_0}),
which means that we have found a closed form solution for the partition
function (\ref{eq:H-eff}) at vanishing applied field.

The Casimir quantities translate into the effective model as follows:
the strip Casimir potential, or strip residual free energy, (\ref{eq:F_res^L-def})
is simply the free energy of the effective model (\ref{eq:H-eff}),
\begin{equation}
\Fsres(T;L,M)=-\log\Zsres=-\log\Tr\ee^{-\mathcal{H}_{\mathrm{eff}}}.\label{eq:fsres}
\end{equation}
By the definition (\ref{eq:dFC^L_a}), the differential Casimir force
per surface area $M$ is given by 
\begin{align}
\FC_{\strip}(T;L,M) & =-\frac{1}{M}\frac{\partial}{\partial L}\Fsres(T;L,M)\nonumber \\
 & =\frac{1}{M}\frac{\partial}{\partial L}\log\Tr\ee^{-\mathcal{H}_{\mathrm{eff}}}\nonumber \\
 & =-\frac{1}{M}\Big\langle\sum_{\mu=1}^{M}\dom{\gamma}_{\mu}s_{\mu}\Big\rangle_{\mathrm{eff}}\Def-m_{\mathrm{eff}}(L),
\end{align}
and is therefore identical to minus the field dependent magnetization
per spin of the effective model in an antiparallel magnetic field
of strength $L$. 

From this mapping, one could conclude that the residual determinant
$\Zsres$ cannot be factorized into a product for arbitrary $L$,
as this would imply an exact solution of a spin system with long range
frustrated interactions in a magnetic field. However, the couplings
(\ref{eq:K_munu}) are products of symmetric functions of the $c_{\mu}$,
which might be utilized to find a factorization. In the finite-size
scaling limit $L,M\to\infty$, $T\to\Tc$, at fixed temperature scaling
variable $x\propto(T/\Tc-1)L$ and aspect ratio $\rho$, such a factorization
indeed exists at least at the critical point $\Tc$. In this limit,
the residual determinant (\ref{eq:Zsres}) can be written in terms
of the Dedekind eta function \cite{Hucht16b}, confirming a result
from conformal field theory \cite{KlebanVassileva91}.

\section{Conclusions}

We have calculated the partition function of the two-dimensional anisotropic
square lattice Ising model on a $L\times M$ rectangle with open boundary
conditions. The final expression (\ref{eq:Z-final-tilde-all}) involves
$M$ eigenvalues $\dom{\lambda}_{\mu}$ of a $M\times M$ transfer
matrix, represented as zeroes of its characteristic polynomial (\ref{eq:CP(phi)}).
The remaining residual part (\ref{eq:Zsres}) is reduced to the determinant
of a $\Mh\times\Mh$ matrix, for which we could not find a closed
product representation (see also \cite{FredMathoverflow16}). 

An analogous calculation, with similar result (\ref{eq:Z-final-tilde-all}),
can be done for arbitrary coupling distributions $z_{\ell<L,m}=z_{m}$,
$t_{\ell,m}=t_{m}$, as long as the involved transfer matrix $\TM 2$,
equation~(\ref{eq:T_2}), is independent of $\ell$. The characteristic
polynomial, eigenvalues and eigenvectors of $\TM{}$, equation~(\ref{eq:T}),
will however be more complicated. On the other hand, we can return
to cylinder geometry with periodic or antiperiodic boundary conditions,
$z_{\ell<L,m}=z$, $t_{\ell,m<M}=t$, $t_{\ell,M}=t^{\pm1}$, in which
case the characteristic polynomial (\ref{eq:CP(phi)}) simply becomes
$P_{M}^{(\mathrm{pbc})}(\varphi)=\cos\!\left(\frac{M\varphi}{2}\right)$
or $P_{M}^{(\mathrm{apbc})}(\varphi)=\sin\!\left(\frac{M\varphi}{2}\right)$
independent of temperature, which greatly simplifies the calculations. 

The intermediate result (\ref{eq:Z_arbitrary}) gives the exact partition
function of the Ising model with arbitrary couplings $K_{\ell,m}^{\L}$
and $K_{\ell,m}^{\M}$ on the cylinder in terms of a product of very
simple $2\times2$ block transfer matrices with $M\times M$ blocks.
This representation can be used to, e.\,g., investigate diluted systems,
or to exactly determine the critical Casimir potential and force between
extended particles on the lattice, as introduced in \cite{HobrechtHucht14,HobrechtHucht15a}.
Due to the reduction to $2M\times2M$ matrices, numerically exact
calculations are possible for large systems up to $M\approx1000$
and arbitrary $L$ on actual personal computers. However, depending
on the actual coupling configuration it might be necessary to use
extended numerical precision.

Finally, we presented an exact mapping of the residual part $\Zsres$
of the partition function onto an effective spin system with long-range
frustrated interactions in an external magnetic field of strength
$L$. This model might serve as starting point for further investigations.

The finite-size scaling limit of the considered model, as well as
results for the Casimir potential and Casimir force scaling functions,
will be published in the second part of this work \cite{Hucht16b}.
\begin{acknowledgments}
The author thanks Hendrik Hobrecht, Felix M. Schmidt and Jesper Jacobsen
for helpful discussions and inspirations, as well as Rodney J. Baxter
for bringing Ref.~\cite{Baxter10a} to my attention. I also thank
my wife Kristine for her great patience during the preparation of
this manuscript. This work was partially supported by the Deutsche
Forschungsgemeinschaft through Grant HU~2303/1-1.
\end{acknowledgments}

\appendix

\section{Product formulas for free energy contributions \label{apx:Product-formulas}}

In this appendix we will give, without derivation, the product formulas
for the singular parts of the free energies $\fb$, $\fs$ and $\fc$
above and below $\Tc$ for the isotropic Ising model, where $K=K^{\L}=K^{\M}$
and $z=t^{*}$. The calculation is done similar to \cite{VernierJacobsen12}:
using the finite lattice method \cite{Enting78} we generate the high-
and low-temperature series expansion of the free energies up to some
finite order and rewrite the series in terms of the natural variable
$q$ \cite{BaxterSykesWatts75} using the inverse Euler transform
\cite{MathWorld-EulerTransform}. Interestingly, both the finite lattice
method and the inverse Euler transform are based on the Möbius inversion
formula from elementary number theory \cite{Hardy79}. The resulting
infinite product in $q$ has a periodic structure 
\begin{equation}
\prod_{k=1}^{\infty}(1-q^{k})^{c_{0,k}+c_{1,k}k},\label{eq:q-Product}
\end{equation}
i.\,e., the coefficients $c_{0,k}$ and $c_{1,k}$ are oscillating
sequences, with period $p\in\{4,8,16\}$, which can be identified.
These sequences are then conjectured to continue to $k\to\infty$.
First we recall the results of Vernier \& Jacobsen \cite{VernierJacobsen12}
obtained for temperatures below $\Tc$.

Infinite products like (\ref{eq:q-Product}) can be written in many
different ways. For the sake of clarity we first introduce a simple
notation for such periodic products: we define the function 
\begin{equation}
\QP(\mat C|q)\Def\prod_{k=1}^{\infty}(1-q^{k})^{c_{k}},\label{eq:Pi-def}
\end{equation}
where the $(m+1)\times p$ coefficient matrix $\mat C$ defines the
$m^{\mathrm{th}}$-order polynomials
\begin{equation}
c_{k}=\sum_{j=0}^{m}\mat C_{j,k\,\mathrm{mod}\,p}\,k^{j}.\label{eq:C-def}
\end{equation}
\begin{figure}
\begin{centering}
\includegraphics[width=0.75\columnwidth]{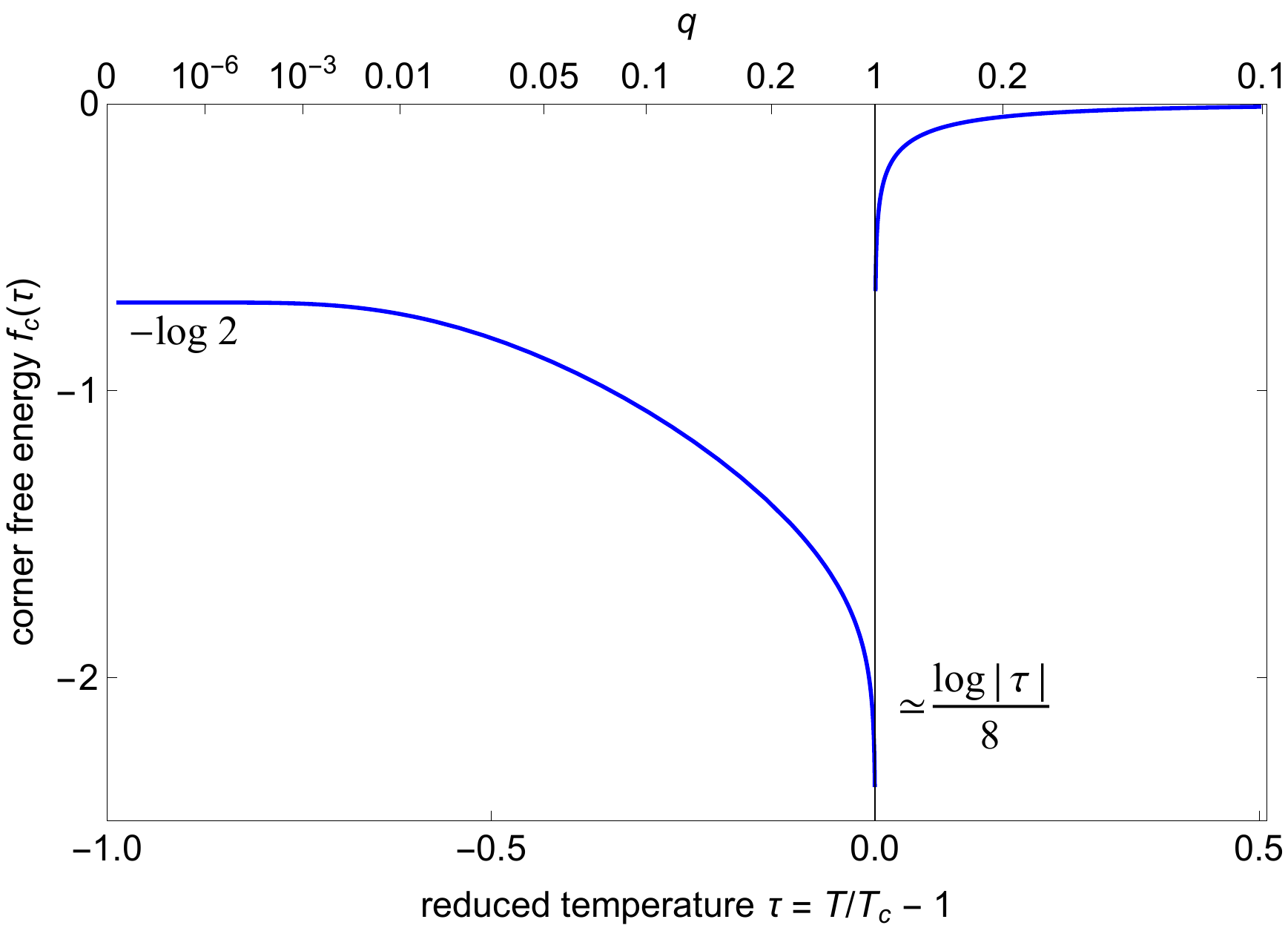}
\par\end{centering}
\caption{Corner free energy $\protect\fc$ vs. reduced temperature of the two-dimensional
Ising model. The corresponding values of the natural variable $q$
are shown at the upper frame. \label{fig:Corner-free-energy}}
\end{figure}
With this definition we first rewrite the results of Vernier \& Jacobsen:
the natural low temperature variable $q$ fulfills \cite[equation (48)]{VernierJacobsen12}
\begin{align}
t^{<} & =\sqrt{q}\,\QP\big(\begin{array}{cccccccc}
0 & 1 & 0 & -1 & 0 & -1 & 0 & 1\end{array}|\,q\big)\label{eq:q-below}\\
 & =\sqrt{q}\frac{(q;q^{8})_{\infty}(q^{7};q^{8})_{\infty}}{(q^{3};q^{8})_{\infty}(q^{5};q^{8})_{\infty}}=\sqrt{q}\frac{(1-q^{1})(1-q^{7})(1-q^{9})(1-q^{15})\cdots}{(1-q^{3})(1-q^{5})(1-q^{11})(1-q^{13})\cdots},\nonumber 
\end{align}
where $t^{<}=\ee^{-2K^{<}}$, and $(a;q)_{\infty}$ denotes the $q$-Pochhammer
symbol. Then, the singular bulk, surface and corner\footnote{Erratum: the constant $-\log2$ in $\fc^{<}$ should be removed, see
\cite{Hucht16b} for details. } free energies become \cite[equation (49)]{VernierJacobsen12} \bS
\begin{align}
e^{-f_{\mathrm{b,sing}}^{<}} & =\frac{1}{\sqrt{q}}\QP\left(\left.\begin{array}{cccccccc}
0 & 0 & -1 & 0 & 2 & 0 & -1 & 0\\
0 & -1 & 0 & 1 & 0 & -1 & 0 & 1
\end{array}\right|q\right),\\
e^{-f_{\mathrm{s,sing}}^{<}} & =\frac{1}{2}\QP\left(\left.\begin{array}{cccccccc}
0 & \frac{3}{4} & -1 & -\frac{3}{4} & 2 & -\frac{3}{4} & -1 & \frac{3}{4}\\
0 & \frac{1}{4} & 0 & \frac{1}{4} & 0 & -\frac{1}{4} & 0 & -\frac{1}{4}
\end{array}\right|q\right)\QP\left(\left.\begin{array}{cccccccc}
0 & -\frac{1}{2} & 0 & \frac{1}{2} & 0 & \frac{1}{2} & 0 & -\frac{1}{2}\\
0 & -\frac{1}{2} & 0 & -\frac{1}{2} & 0 & \frac{1}{2} & 0 & \frac{1}{2}
\end{array}\right|\sqrt{q}\right),\\
e^{-\fc^{<}} & =2\QP\left(\left.\begin{array}{cccccccc}
0 & -2 & 3 & -2 & -1 & -2 & 3 & -2\\
0 & -2 & \frac{1}{2} & 2 & 0 & -2 & -\frac{1}{2} & 2
\end{array}\right|q\right),
\end{align}
\eS{}where the regular part of $\fc$ is zero, while $f_{\mathrm{b,reg}}=-\log[2(1+z^{2})/(1-z^{2})]$
and $f_{\mathrm{s,reg}}=-\frac{1}{4}\log(1-z^{2})$. Doing the same
analysis in the paramagnetic phase we first identify the high temperature
variable $z^{>}=\tanh K^{>}$ by duality \cite{BaxterSykesWatts75},
such that
\begin{equation}
z^{>}=\sqrt{q}\,\QP\big(\begin{array}{cccccccc}
0 & 1 & 0 & -1 & 0 & -1 & 0 & 1\end{array}|\,q\big)\label{eq:q-above}
\end{equation}
has the same product representation as (\ref{eq:q-below}). Then we
find the infinite products \bS
\begin{align}
e^{-f_{\mathrm{b,sing}}^{>}} & =\QP\left(\left.\begin{array}{cccccccc}
0 & 2 & -4 & 2 & 0 & 2 & -4 & 2\\
0 & -1 & 0 & 1 & 0 & -1 & 0 & 1
\end{array}\right|q\right)=\QP\left(\left.\begin{array}{cccc}
0 & 2 & -4 & 2\\
0 & -1 & 0 & 1
\end{array}\right|q\right),\\
e^{-f_{\mathrm{s,sing}}^{>}} & =\QP\left(\left.\begin{array}{cccccccc}
0 & \frac{1}{4} & 1 & -\frac{1}{4} & -2 & -\frac{1}{4} & 1 & \frac{1}{4}\\
0 & -\frac{1}{4} & 0 & -\frac{1}{4} & 0 & \frac{1}{4} & 0 & \frac{1}{4}
\end{array}\right|q\right),\\
e^{-\fc^{>}} & =\QP\left(\left.\begin{array}{cccccccc}
0 & 0 & 0 & 0 & -3 & 0 & 0 & 0\\
0 & 0 & -\frac{1}{2} & 0 & 0 & 0 & \frac{1}{2} & 0
\end{array}\right|q\right)=\QP\left(\left.\begin{array}{cccc}
0 & 0 & -3 & 0\\
0 & -1 & 0 & 1
\end{array}\right|q^{2}\right)\\
 & =\prod_{k=0}^{\infty}\frac{1}{(1-q^{2(4k+2)})^{3}}\frac{(1-q^{2(4k+3)})^{4k+3}}{(1-q^{2(4k+1)})^{4k+1}}.
\end{align}
\eS{}Note that the period of all three products above $\Tc$ is half
of the period below $\Tc$ ($e^{-f_{\mathrm{s,sing}}^{<}}$ can be
written as a single product in $\sqrt{q}$, with period 16). The second
product in $e^{-f_{\mathrm{s,sing}}^{<}}$ is interpreted as the additional
contribution from the surface tension. The corner free energy $\fc^{>}$
can be written as a function of $q^{2}$, because $c_{k}\neq0$ only
for even numbers $k$. Finally, we show the corner free energy $\fc$
in figure~\ref{fig:Corner-free-energy}. For $T\to0$, $\fc\to-\log2$,\footnote{Erratum: $\fc(T\to0)\to0$, see \cite{Hucht16b} for details. }
while for $T\to\Tc$ we find a logarithmic divergence from both sides,
with different amplitudes. A detailed discussion of the critical region
will be presented in \cite{Hucht16b}.

\bibliographystyle{unsrt}
\addcontentsline{toc}{section}{\refname}\bibliography{Physik}

\end{document}